\documentclass{aa}

\usepackage{graphicx}
\usepackage{txfonts}
\usepackage{amsmath}

\usepackage{natbib}
\begin{document} 

  \title{\emph{Gaia} Data Release 2. Cross-match with external catalogues }
  
   \subtitle{Algorithms and results}

   \author{P.M. Marrese \inst{1,2}\fnmsep\thanks{ \email{paola.marrese@ssdc.asi.it}}
       \and S. Marinoni \inst{1,2}
       \and M. Fabrizio \inst{1,2}
       \and G. Altavilla \inst{1,2}
   }
   \institute{ INAF-Osservatorio Astronomico di Roma, Via Frascati 33, I-00078, Monte Porzio Catone (Roma), Italy
   \and Space Science Data Center - ASI, via del Politecnico SNC, I-00133 Roma, Italy\\ }
       
   \date{Received ; accepted}

 
  \abstract
  {Although the \emph{Gaia} catalogue on its own is a very powerful tool, it is the combination of this high-accuracy archive with other archives that 
   will truly open up amazing possibilities for astronomical research. The advanced interoperation of archives is based on cross-matching, leaving 
   the user with the feeling of working with one single data archive. The data retrieval should work not only across data archives but also across
   wavelength domains. 
   The first step for a seamless access to the data is the computation of the cross-match between \emph{Gaia} and external surveys.}
  {We describe the adopted algorithms and results of the pre-computed cross-match of the  \emph{Gaia} Data Release 2 (DR2) catalogue with dense surveys (Pan-STARRS1~DR1, 2MASS, SDSS~DR9, GSC~2.3, URAT-1, allWISE, PPMXL, and APASS~DR9) and sparse catalogues (Hipparcos2, Tycho-2, and RAVE~5).}
  {A new algorithm is developed specifically for sparse catalogues. Improvements and changes with respect to the algorithm adopted for DR1 are described in detail.} 
  {The outputs of the cross-match are part of the official \emph{Gaia}~DR2 catalogue. The global analysis of the cross-match  results is also presented. }
   {}


   \maketitle
%
\section{Introduction}\label{intro}
The \emph{Gaia} satellite allows determining high-accuracy positions for  $\sim$1.7 billion sources and parallaxes and proper motions for  $\sim$1.3 billion sources  
observed all-sky down to magnitude G$\sim$20.7. Compared to the first intermediate \emph{Gaia} Data Release (DR1, see \citealt{Brown2016} for  a summary of the astrometric, photometric, and survey properties, and \citealt{Prusti2016}  for the scientific goals of the mission), the second intermediate \emph{Gaia} Data Release \citep{Brown2018} provides  48\% additional sources, parallaxes, and proper motions with an unprecedented accuracy for 77\% of all observed sources, which are complemented by a precise and homogeneous multi-band  photometry and a large radial velocity survey for more than  7\,000\,000 sources with G magnitude in the $4-13$ range. Astrophysical parameters for $\sim$160 million sources, data on more than 500\,000 variable stars, and $\sim$14\,000 solar system objects are also available in DR2\footnote{A more exhaustive overview of the mission and DR2 details can be found at \url{ https://www.cosmos.esa.int/web/gaia/dr2-papers}}.

The main goal of adding a pre-computed cross-match to \emph{Gaia}~DR2 data is complementing \emph{Gaia} with existing astrophysical quantities (that are widely used by the scientific
community). This allows the full exploitation of the scientific potential of \emph{Gaia} .

The general principles of the adopted cross-match algorithm are given and discussed in \citealt{Marrese2017} (hereafter Paper~I). We here briefly recall that any cross-match algorithm is a trade-off between multiple requisites, and a fraction of mismatched and/or missed objects is always present. Our aim is to define and implement a cross-match algorithm that on one hand should be general enough to be exploited for different scientific cases, and on the other  should have complete results that can later be filtered to better fullfil a specific scientific problem. We tried to find a reasonable compromise between the completeness and correctness requirements, which implies that we needed to avoid adding too many spurious matches.

In Sections~\ref{sec:gen} and \ref{sec:details} we describe the general principle and the details of the cross-match algorithms defined for  \emph{Gaia}~DR2, respectively. Section~\ref{sec:extcat} contains the list of the external catalogues that we matched with \emph{Gaia}~DR2 data and a short description for the newly added catalogues, together with some  issues or caveats that are relevant to the cross-match. In Sections~\ref{sec:output} and \ref{sec:results} we describe and discuss the cross-match output content and the results.
Finally, Appendix~\ref{sec:app} contains a discussion of the effective angular resolution of external catalogues and its influence on the cross-match.

\section{\emph{Gaia} pre-computed cross-match: general principles\label{sec:gen}}

\begin{figure}
   \centering
   \includegraphics[width=0.92\linewidth]{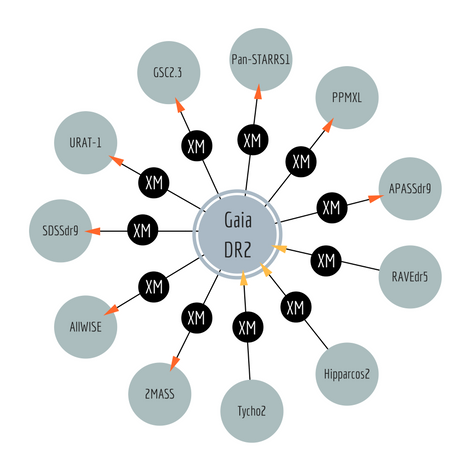}
   \caption{\emph{Gaia} DR2 cross-match schema: for large dense surveys (orange arrows), \emph{Gaia} is the leading catalogue, while for sparse catalogues 
   (yellow arrows), the external catalogue leads. }
   \label{figure:XMscheme}
   \end{figure}

Following the same approach as in Paper~I, we define the cross-match algorithm according to the scientific problem we are faced with. 
Since the cross-match results with external catalogues are part of the official \emph{Gaia} DR2 and are integrated in the \emph{Gaia} catalogue access environments, it is fundamental to match \emph{Gaia} with each external survey separately and independently, in a consistent and homogeneous manner.  We therefore created links between different surveys through the \emph{Gaia} catalogue, which is all-sky and has the highest angular resolution. \emph{Gaia} is thus at the centre of our cross-match schema, as depicted in Figure~\ref{figure:XMscheme}.
For multi-catalogue searches, the catalogue specific matches to \emph{Gaia} that are common to different surveys can be selected using multiple joins.

The external catalogues to be matched with \emph{Gaia} DR2 are all obtained in the optical/near-IR wavelength region (with the exception of allWISE, which extends in the medium-IR domain), are general surveys not restricted to a specific class of objects, and have an angular resolution lower than \emph{Gaia}, as was the case for \emph{Gaia} DR1. 
However, in contrast to  the case of the cross-match of \emph{Gaia} DR1, the external catalogues to be matched with \emph{Gaia} DR2 are not sufficiently homogeneous among themselves for the exact same algorithm to be used for all of them. We therefore broadly separated the external catalogues into two different groups: large dense surveys, and sparse catalogues, and we defined two slightly different algorithms for the two groups. 
External catalogues are here defined as dense surveys when it is possible to define a precise (i.e. based on a reasonable number of objects) and accurate (i.e. local) density around the majority of their objects.
The two algorithms we define are not symmetric, and for the dense surveys, we use \emph{Gaia} as the leading catalogue, while for sparse catalogues, we use \emph{Gaia}  as the second catalogue.

The cross-match algorithms we use in DR2 are quite similar to the algorithm that was successfully used in DR1, however we could take advantage of the enormous increase in the number of sources with proper motions 
and parallaxes with respect to \emph{Gaia} DR1, and we ameliorated the algorithm in many respects:  \emph{a)} use of the full five-parameter covariance matrix, \emph{b)} improved density definition, \emph{c)} source-by-source definition of the initial search radius,  which allows matching high proper motion stars, and \emph{d)} definition of the proper motion threshold to be used for \emph{Gaia} sources with no proper motions based on a trade-off between completeness and correctness.

Similarly to what was done for DR1, in the \emph{Gaia} DR2 cross-match algorithms, we have not defined any special treatment for binary stars so far.
The binary stars that may represent a problem for the cross-match are physically related sources with an additional motion
that is due to multiplicity, which can displace their positions enough to prevent them from matching. 
As a general principle, when we knew of an effect that influences astrometry (and thus the cross-match results) but there was no indication in \emph{Gaia} data how strongly this would affect a specific source, we added a systematic to all affected sources, as we did when we broadened the position errors of  \emph{Gaia} sources without proper motions (see Subsection~\ref{sub:epochdiff}). However, when an effect influences the astrometry of a specific subsample of sources (such as binaries), but there is no information on which sources and how strong the influence is in the  \emph{Gaia} data, we assumed a more cautious attitude and only added a caveat stating that the effect was not taken into account.

We repeat here some basic definitions that are still valid in DR2, but can also be found in Paper~I. A good neighbour for a given object in the leading catalogue is a nearby object in the second catalogue whose 
position is compatible within position errors with the target. We assume that when a good neighbour is found, it is the counterpart. 
When more than one good neighbour is found, the best neighbour (i.e. the most probable counterpart according to the figure of merit we define, see Section~\ref{sec:details}) is chosen among the good neighbours.
Also for \emph{Gaia} DR2, we produced two separate cross-match outputs: a BestNeighbour table, which lists the leading catalogue matched objects with their best neighbour, and a Neighbourhood table, which includes all good neighbours for each matched object (see Section~\ref{sec:output} for a detailed output description).

For dense surveys, the higher angular resolution of \emph{Gaia}  requires a many-to-one algorithm: therefore the algorithm we used is not symmetric and more than one \emph{Gaia} object can have the same best neighbour in a given dense survey.
Two or more \emph{Gaia} objects with the same best neighbour are called mates. True mates are objects that are resolved by \emph{Gaia}, but are not resolved by the external survey. 
For sparse catalogues (such as Hipparcos2, Tycho-2, and RAVE~5), where the external catalogue is the leading catalogue, a one-to-one match is forced and mates are not allowed. Additional good neighbours in \emph{Gaia}  for each sparse catalogue source can be found in the Neighbourhood output table.

The cross-match algorithms used for \emph{Gaia} DR2 are positional and evaluate the second catalogue environment, like for DR1. However, for DR2, we  exploit the full  five-parameters covariance matrix calculated for the \emph{Gaia} astrometric solution (\citealt{Lindegren2018}, \citealt{Mignard2018}) when it is available (i.e. ~77\% of sources). 

\subsection{Accounting for epoch differences\label{sub:epochdiff}}
  
Cross-match algorithms are based on the comparison of source positions in different surveys. Surveys can have been obtained at different epochs, which can be decades apart, and sources often move appreciably in the meantime, therefore it is important to take the source motion into account.
 \begin{figure}[!b]
   \centering
   \includegraphics[width=0.96\linewidth]{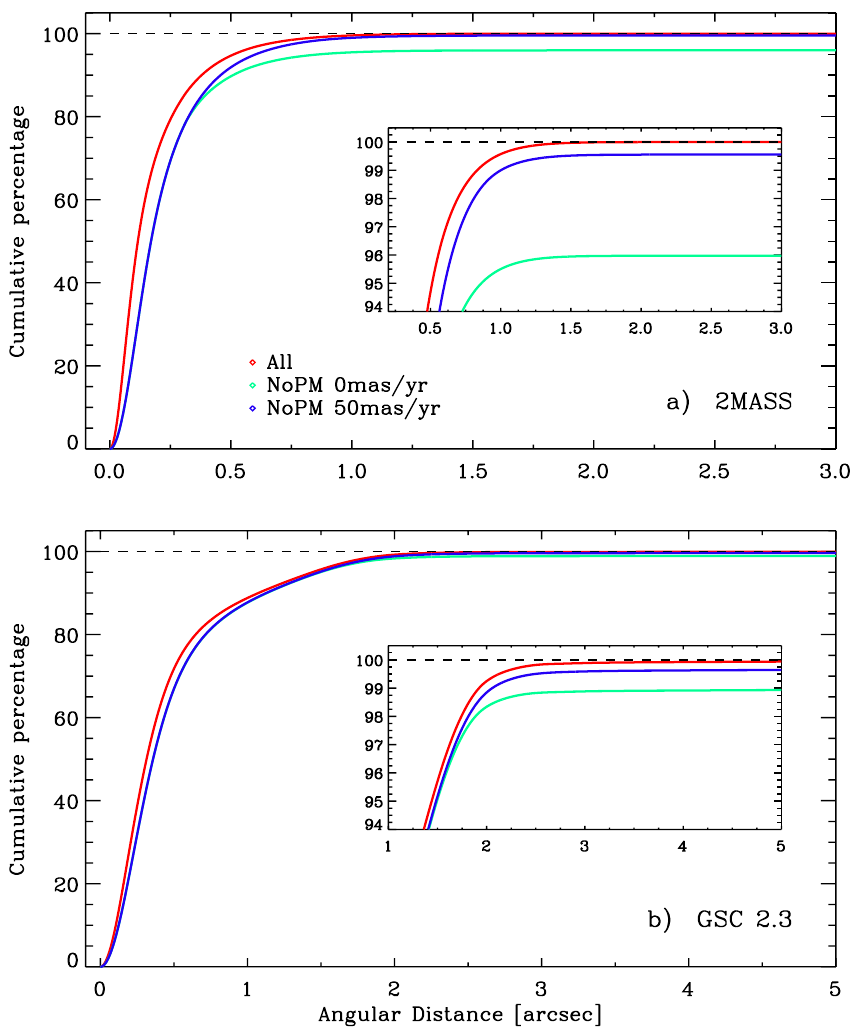}
   \caption{Cumulative distribution of the angular distance of correct best matches obtained with the broadening method using two different proper motion thresholds  (50~mas/yr and 0~mas/yr) when compared to the proper motion propagation of positions method (red curve indicated by \emph{All} label).}
              \label{figure:PMthresh}
    \end{figure}
\begin{table*}
\tiny
\caption{Comparison of the correct best matches obtained with the broadening method using two different proper motion thresholds (and the 
proper motion propagation of positions method as baseline).}   
\centering          
\label{table:PMthresh}      
\begin{tabular}{ l  r  r  r   r  r  r }     
\hline
Catalogue       & $N$ Best matches & Proper motion& $N$ Best matches & \% Best matches & $N$ Correct best matches &\% Correct best matches \\
                       & pos prop\tablefootmark{a} method\tablefootmark{b}& threshold (mas/yr) & broad method\tablefootmark{b}  & broad method\tablefootmark{b} & broad method\tablefootmark{b}  & broad method\tablefootmark{b} \\ \hline 
2MASS PSC   & 424\,265\,005\phantom{000}    &50  &432\,794\,791 & 102.01 & 422\,382\,563 &  99.56\\      
                       & 424\,265\,005\phantom{000}    &0    &408\,841\,264 & 96.36  & 407\,427\,704 &  96.03 \\
GSC~2.3          & 727\,460\,368\phantom{000}    & 50 & 731\,416\,596 & 100.54 & 725\,395\,647 & 99.72\\
                       & 727\,460\,368\phantom{000}    & 0   & 725\,035\,070 & 99.67 & 723\,152\,510 & 99.41\\
\hline          
\end{tabular}
\tablefoot{
   \tablefoottext{a}{pos prop: proper motion propagation of positions method.}
   \tablefoottext{b}{We list here the number of distinct external catalogue sources that matched with a  \emph{Gaia} source with a five-parameter astrometric solution.}
    }
\end{table*}
\begin{table*}[t]
\small
\caption{Fraction of objects with a given number of nearby sources that was used to evaluate the local surface density together with the radius within which the nearby sources are found (see Subsection~\ref{subsec:environment}).}   
\centering          
\label{table:Density}      
\begin{tabular}{ l  r  r  r   r  r  r}     
\hline
Catalogue                 & $Radius_{\mathrm{max}}$  & \% sources      & \% sources      & \% sources      & \% sources      & \% sources           \\ \hline 
                                 &         (arcsec)                      & $N_{stars}<10$ &$10<=N_{stars}<30$&$30<=N_{stars}<50$& $50<=N_{stars}<100$ & $N_{stars}=100$ \\ 
\emph{Gaia} DR2     & 300 &0.000001 &0.0002     & 0.11        & 7.24 & 92.65\\
Pan-STARRS1~DR1       &  120 &0.0004     &0.0326       &0.21    &      30.58  &69.18\\
GSC~2.3                   & 480 &0.0001     & 0.000007& 0.0003    & 0.22 &99.78\\
PPMXL                    & 480 &0.000001  & 0.000006& 0.000026& 0.51 &99.48\\
SDSS DR9              & 600 &0.000006 & 0.00004   & 0.00005  & 0.006&99.99\\
URAT-1                   & 480 &0.001       & 0.05          & 0.156     & 13.31 &86.48\\
2MASS PSC           & 600 & 0            & 0               & 0.007      & 2.52    &97.48 \\
allWISE                   & 480 &0& 0 &0.000008&0.0068&99.99 \\
APASS DR9           &600&0.0061&1.52&6.27&28.81& 63.40 \\
\hline          
\end{tabular}
\end{table*}

In order to do so, we moved the \emph{Gaia} objects to the individual epoch of the possible matches in the external catalogues using
the algorithm provided in the Hipparcos and Tycho Catalogue documentation (\citealt{Hipparcos}). 
While this algorithm requires the use of all six parameters, $\alpha$ (Right Ascension), $\delta$ (Declination), $\pi$ (parallax), $\mu_{\alpha*}$ (proper motion in $\alpha\cos\delta$),  
$\mu_{\delta}$ (proper motion in $\delta$), and $V_{\mathrm{R}}$ (radial velocity), $V_{\mathrm{R}}$ is not included in the published astrometric solution  in  \emph{Gaia} DR2. Nonetheless, according to  
\citet{Lindegren2018}, and in particular their Section~3, $V_{\mathrm{R}}$  is relevant only for very few sources (53). 

For the fraction of \emph{Gaia} sources for which only a position (i.e. 2 parameters) astrometric solution is available,  we applied the broadening method described in Paper~I for the sake of completeness. We thus defined a proper motion threshold that is common to all sources and all external catalogues for homogeneity and consistency reasons. In DR1, the adopted threshold (200~mas/yr) was chosen by evaluating the distribution of known high proper motion stars.
Instead, while we are aware that the peak of the total proper motion distribution for Gaia sources is $\sim$6~mas/yr, and with the aim of also recovering high proper motion stars in the subsample of \emph{Gaia} sources with two-parameter astrometric solutions, we decided for DR2 to derive  from the data which was the most appropriate proper motion threshold to use.
We therefore considered the subsample of  \emph{Gaia} sources with a five-parameter astrometric solution, and we compared the cross-match results obtained using the position propagation method on one hand and the broadening method on the other. In this evaluation, we assumed  \emph{a)} that the subsample of \emph{Gaia} sources without available proper motions has the same proper motion distribution as the subsample with measured proper motions,  and \emph{b)} that the result obtained using the position propagation is correct. We conducted different tests with the broadening method in order to determine a proper motion threshold that allowed maximising the number of correctly recovered matches and minimising the addition of spurious matches. The tests were performed on all catalogues using different thresholds. The number of sources recovered in the cross-match output for a given external catalogue depends on the combination of the typical epoch difference between the external catalogue and  \emph{Gaia}~DR2 and the typical size of the position errors of the external catalogue. The larger the epoch difference and the smaller the position errors, the larger  the number of recovered sources and thus the more relevant the position error broadening. On the other hand, the denser the external catalogue, the larger the number of added spurious matches.

Table~\ref{table:PMthresh} and Figure~\ref{figure:PMthresh} illustrate the method we used and show the comparison of the cross-match results for 2MASS PSC and GSC~2.3 between the position propagation method, the broadening method with the adopted 50 mas/yr threshold, and the method without position propagation (i.e. broadening threshold 0~mas/yr). In order to describe how the position error broadening method works, we chose two catalogues: 2MASS, for which the method gives a good improvement
in the number of matched sources, and GSC~2.3, for which the improvement is less relevant.
In the case of 2MASS, the typical epoch difference is $\sim$15 years, which combined with a typical  \emph{Gaia} total proper motion of 6 mas/yr, implies a $\sim$0.09 arcsec displacement. 
This displacement must be compared with the 2MASS position errors, which for most of the sources, are smaller than 0.1 arcsec. In the case of GSC~2.3, instead, the typical epoch difference with 
\emph{Gaia} is $\sim$25 years, which implies a displacement due to proper motions of about 0.15 arcsec. This displacement is small compared with the 0.3$-$0.4 arcsec values of the typical GSC~2.3 position errors.

Therefore, the adopted proper motion threshold for DR2 is 50 mas/yr. This is our best compromise between completeness and the quantity of spurious matches added to the cross-match.

\subsection{Environment\label{subsec:environment}}
As discussed in Paper~I, the cross-match is not only a source-to-source but also a local problem, thus the figure of merit used to evaluate the good neighbours 
{and to choose the best neighbour among them} should also take into account the local surface density of the second catalogue. 
The density is thus included in the adopted figure of merit (FoM, see Subsection~\ref{subsec:fom}), and its precision (which depends on the number 
of sources used to obtain it) has an important influence on the FoM precision. 
Ideally, $\sim$100  sources are required to evaluate the FoM with a good precision, 
while $\sim$30 sources are still acceptable. The radius needed to obtain the minimum number of 
sources is instead a measure of the accuracy of the density and consequently of the FoM. 
A more local determination is indeed more accurate, especially in dense fields, where there 
are density variations on small scales and where the FoM is more important as it is used 
to select the best neighbour among an higher number of good neighbours.

For \emph{Gaia} DR2, the local density was pre-calculated around each second catalogue source and was fed to the cross-match algorithm. We used a \emph{K}-nearest method that aims to determine the  radius at which the 100th nearby source is found. We also set a maximum radius to search for nearby sources that depends on the catalogue number of sources weighted by its sky coverage.  
The reason we defined a maximum radius is that we consider an accurate (i.e. local) density more important than a precise density, but computation performances were also taken into account.
When the algorithm reached the maximum radius threshold, the corresponding star number was used to calculate the density, even if it was lower than 100. Table~\ref{table:Density} allows determining for each catalogue, including \emph{Gaia}, the fraction of sources with a sub-optimal density determination.

The density determination is improved for DR2 with respect to DR1. However, it is a compromise just like many other details of the cross-match algorithm described in this paper, 
specifically, a compromise between accuracy and precision.
\section{\emph{Gaia} pre-computed cross-match: details\label{sec:details}}
  \begin{figure*}
   \centering
   \includegraphics[width=0.96\linewidth]{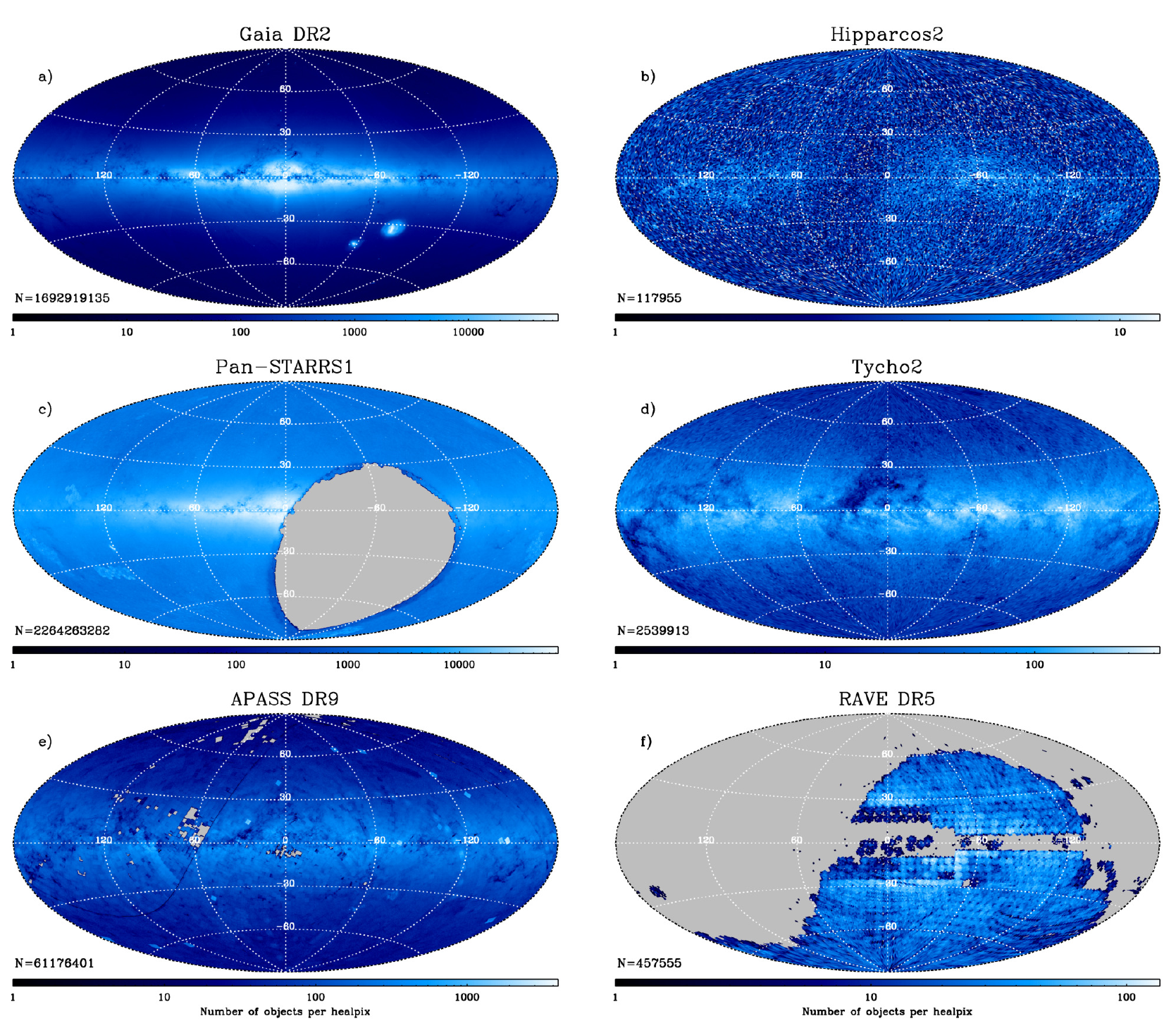}
   \caption{Surface density distribution for \emph{Gaia} DR2 and the new external catalogues 
(see Section~\ref{sec:extcat}) obtained using a HEALPix (Hierarchical Equal Area 
isoLatitude Pixelization, \citealt{gorski}) tessellation with 
resolution $N_{\mathrm{side}}=2^{8}$ for dense surveys and 
$N_{\mathrm{side}}=2^{6}$ for sparse catalogues. 
   In grey we indicate areas that are not covered by the survey. The surface density distribution of external catalogues that were also matched with \emph{Gaia} DR1 can be found in Paper~I.}
              \label{figure:all}
    \end{figure*}
\begin{table*}[t]
\small
\caption{\emph{\textup{Properties of} Gaia} DR2 and external catalogues.}    
\centering          
\label{table:ExtProp}      
\begin{tabular}{ l  r  l  c   l  l  l }     
\hline
Catalogue & $N$ Sources & $PosErr_{\mathrm{max}}$\tablefootmark{a} & Effective resolution & $\Delta Epoch_{\mathrm{max}}$&$SysErr_{\mathrm{max}}$\tablefootmark{b} &Survey type\tablefootmark{c}\\
                                &                       &\multicolumn{1}{c}{(arcsec)}  & (arcsec) &\multicolumn{1}{c}{(yr)}  & (arcsec) & \\ \hline 
\emph{Gaia} DR2    &1\,692\,919\,135&\phantom{222}{0.1}                                  & 0.4\tablefootmark{d}                      &\phantom{2222}  & & \\
Pan-STARRS1~DR1       &2\,264\,263\,282&\phantom{222}1.0                                      & $\sim$1.1                                                       &\phantom{222}18.02 & 0.18 & Dense\\ 
GSC~2.3                   &945\,592\,683  &\phantom{222}1.6                                      & $\sim$2\tablefootmark{f}            &\phantom{222}62.79&0.63 & Dense\\
PPMXL                    &910\,468\,688  &\phantom{222}1.342\tablefootmark{e}       & $\sim$2\tablefootmark{f}            &\phantom{222}15.5 &0.155 & Dense\\ 
SDSS DR9              &469\,029\,929  &\phantom{22}10.0                                       &  $\sim$0.7\tablefootmark{f}            &\phantom{222}16.79 & 0.17 & Dense\\
URAT-1                   &228\,276\,482  &\phantom{222}0.429                                   &  $\sim$2.5\tablefootmark{f}               &\phantom{2222}3.189 &0.03 & Dense\\ 
2MASS PSC           & 470\,992\,970  &\phantom{222}1.21                                    & $\sim$2.5                                                      & \phantom{222}17.29 &0.173 & Dense\\ 
allWISE                   &747\,634\,026   &\phantom{22}35.944                                  & 6.1, 6.8, 7.4,12.0\tablefootmark{g}   &\phantom{2222}5.47 & 0.055 & Dense\\ 
APASS DR9           &61\,176\,401       &\phantom{222}2.359                                 & $\sim $5                                          &\phantom{2222} 3.5  & 0.035  &  Dense\\  
Hipparcos2            & 117\,955            &\phantom{222}0.1684                                & $\sim $0.3                                       & \phantom{222}24.25&0.2425 & Sparse\\
Tycho-2                   &2\,539\,913        &\phantom{222}0.254                                  &  $\sim$0.8                                                    &\phantom{222}24.275&0.2475 & Sparse\\
RAVE~5                  &457\,555\tablefootmark{h}             &\phantom{222}0.6\tablefootmark{i}            &3.5\tablefootmark{l}                       &\phantom{222}15.5\tablefootmark{i}&0.155 & Sparse\\
\hline          
\end{tabular}
\tablefoot{
   \tablefoottext{a}{$PosErr_{\mathrm{max}} = \max [\max(RAerr),\max(DECerr) $].}
   \tablefoottext{b}{$SysErr_{\mathrm{max}} = PM \cdot \Delta Epoch_{max}/5$, see Subsection~\ref{subsec:broad}.}
   \tablefoottext{c}{This classification determines the algorithm used to calculate the cross-match, see Section~\ref{sec:gen}.}
   \tablefoottext{d}{See \citet{Arenou2018}, Subsection~2.2, for an analysis of the effective angular resolution of \emph{Gaia} DR2.}
   \tablefoottext{e}{The maximum of the position error refers to the propagated errors at J2000.0.}
   \tablefoottext{f}{Effective resolution value is our best guess (see Section~\ref{sec:extcat}).}
   \tablefoottext{g}{Angular resolution in the four bands W1,W2,W3, and W4.}
   \tablefoottext{h}{RAVE contains multiple observations of the same source, which are identified with the same RAVEID and a different RAVE\_OBS\_ID. The number reported in this table is the number of RAVE distinct sources.}
   \tablefoottext{i}{Assumed for cross-match calculation, see Section~\ref{sec:details}.}
   \tablefoottext{l}{For RAVE~5, we used roughly half the size in the sky of the fiber diameter of the multi-object spectrograph 6dF that was used to observe the RAVE sources.}
   }
\end{table*}
We recall here the basic details of the cross-match algorithm and outline the differences with DR1.
As described in Paper~I, the cross-match algorithm uses a plane-sweep technique that requires the catalogues to be sorted by declination, implies the definition of an active list of objects in the second catalogue for each leading catalogue object, and allows reading the input data only once, which speeds up the cross-match computation (\citealt{Power2005}, \citealt{Abel2004}, \citealt{Devereux2004}, \citealt{Power2004}). 
We used the same filter and refine technique as in DR1, but the first filter is now defined on an object-by-object basis (i.e. it is different for each target within a given leading catalogue), rather than being fixed for a given pair of leading and second catalogues, and it is only used to select candidate good neighbours and not to calculate the density on-the-fly.
The second filter is used to select good neighbours among the candidates. The selection of the best neighbour among good neighbours is based on the same FoM as described in Paper~I.
A normal distribution for position errors is still assumed, and the position error ellipses are projected on the tangent plane. 

\subsection{Initial search radius (first filter)}
In the following, subscript \emph{L} stands for leading catalogue and subscript \emph{S} stands for second catalogue.
The definition of the initial search radius ($R_{I}$) depends on the position in the cross-match algorithm of the \emph{Gaia} catalogue as leading (i.e. dense surveys cross-match) or second catalogue (i.e. sparse catalogues cross-match).

\noindent
$R_{I}$ is computed around each leading catalogue object as
   \begin{equation}
      R_{I} = H_{\gamma} \cdot PosErr_{\mathrm{L,max}} + \left(\frac{PM \cdot \Delta Epoch_{\mathrm{L,max}}}{1000} \right),
   \end{equation}
\noindent
 where  $H_{\gamma} = 5$ corresponds to a confidence level $\gamma$ of 0.9999994267;
 $PosErr_{\mathrm{L,max}} $ is the combined position error for each L source with the maximum position error in the S catalogue; 
 $\Delta Epoch_{\mathrm{L,max}}$ is the maximum reference epoch difference between the L source and the S catalogue; and      
 $PM$ is the proper motion considered.  The definition of  $PM$ is different in different cases: 
 \begin{itemize}
 \item{proper motion of the L source if  \emph{Gaia} is the leading catalogue and the L source has a five-parameter astrometric solution;}
 \item{proper motion threshold if  \emph{Gaia} is the leading catalogue and the L source has a two-parameter astrometric solution;}
 \item{maximum of the \emph{Gaia} catalogue proper motions if   \emph{Gaia} is the second catalogue.}
 \end{itemize}
 \noindent
The combined position error $PosErr_{\mathrm{L,max}} $ is now defined as
\begin{equation}\begin{split}
PosErr_{\mathrm{L,max}} =~&\max[ RAerr_{L},DECerr_{L}] + \\&\max[ \max(RAerr_{S}),\max(DECerr_{S}) ] 
\end{split},\end{equation}
\noindent
where $RAerr$ and $DECerr$ are the uncertainties in Right Ascension and Declination.
The maximum epoch difference between the L source and the S catalogue being matched is defined as
\begin{equation}\begin{split}
\Delta Epoch&_{\mathrm{L,max}} = \\& \max \Big[~|\max(refEpoch_{L})-\min(refEpoch_{S})|~,\\& |\min(refEpoch_{L})-\max(refEpoch_{S})|~ \Big]
\end{split}.\end{equation}
In the above equations, $R_{I}$ is in arcsec, $PosErr_{\mathrm{L,max}}$ in arcsec, $PM$ is in mas yr$^{-1}$, and $refEpoch$ is in years.

\subsection{Broadening of position errors\label{subsec:broad}}
While a detailed discussion of the broadening method is available in Paper~I, we repeat here for clarity the equations defining the position error broadening:
\begin{equation}\begin{split}
&\sigma_{x_{G'}} = \sigma_{x_{G}} +  SysErr_{x} =  \sigma_{x_{G}} +  PM \cdot \Delta Epoch/5 \\&
\sigma_{y_{G'}} = \sigma_{y_{G}} +  SysErr_{y} =  \sigma_{y_{G}} +  PM \cdot \Delta Epoch/5
\end{split}
,\end{equation}
\noindent
where G stands for \emph{Gaia}. In DR2 we always broadened the \emph{Gaia} position errors when a five-parameter astrometric solution was not available, regardless of whether \emph{Gaia} was the leading or second catalogue. 

\subsection{Good neighbour selection (second filter)}
In order to define the second filter, it was necessary to convolve the leading and second catalogue position errors. We refer to \citet{pineau} and to Paper~I for a detailed definition and derivation of the position error convolution ellipse.

The second filter is based on the Mahalanobis normalised distance $K_{\gamma}$ (see Equation~9 in Paper~I). $K^{2}_{\gamma}$ has
a $\chi^2$ distribution with two degrees of freedom, and its adopted value corresponds to a value of the confidence level $\gamma$ of 0.9999994267, which in 1D is equivalent to 5$\sigma$. Good neighbours are defined as neighbours that fall within the ellipse defined by the confidence level $\gamma$.
The second filter is thus defined as
\begin{equation}\label{eq:secondfilter}
\frac{d}{\sigma_{x_{C}} \sqrt{ 1 - \rho_{C}^{2}} } \leq K_{\gamma},\\
\end{equation}
where $d$ is the angular distance, $\sigma_{x_{C}}$ is the convolution ellipse error in the direction from the leading catalogue object to the possible counterpart, and
$ \rho_{C}$ is the correlation between $\sigma_{x_{C}}$ and $\sigma_{y_{C}}$.
The high-confidence level was chosen in order to improve the completeness of the cross-match. 

\subsection{Best neighbour selection: figure of merit\label{subsec:fom}}

The FoM we used to select the best neighbour among the good neighbours evaluates the ratio between two opposite models/hypotheses:
the counterpart candidate is a match or it is found by chance. The FoM depends on the angular distance and the position errors, on the epoch difference, 
and on the local surface density of the second catalogue. 
For each of the good neighbours, we computed the FoM and the derived score, described in detail in Paper~I.
The score is listed in the Neighbourhood output table. The best neighbour is defined as the good neighbour with the highest score value.

\section{External catalogue characteristics\label{sec:extcat}}

Following is the list of external catalogues that were cross-matched with the \emph{Gaia} DR2 catalogue and had already been matched with DR1:
\begin{itemize}
\item{GSC~2.3 (\citealt{GSC2.3})}
\item{PPMXL (\citealt{PPMX,PPMXL})}
\item{SDSS DR9 primary objects (\citealt {SDSS9,SDSS12})}
\item{URAT-1 (\citealt{URAT1}) }
\item{2MASS PSC (\citealt{2MASS})}
\item{allWISE (\citealt{WISE,allWISE})}
\end{itemize}
Following is the list of the new external catalogues that were cross-matched with \emph{Gaia} DR2:
\begin{itemize}
\item{Pan-STARRS1~DR1 (\citealt{panstarrs1,panstarrs1b,panstarrs1c,panstarrs1d,panstarrs1e,panstarrs1f})}
\item{APASS DR9 (\citealt{apass9})}
\item{Hipparcos2 (\citealt{Hipparcos,Hipparcos2})}
\item{Tycho-2 (\citealt{Hipparcos,Tycho2})}
\item{RAVE~5 (\citealt{rave5,RAVEON})}
\end{itemize}
The main properties to consider when matching the external catalogues with \emph{Gaia} are 
\emph{a)}~the effective angular resolution, \emph{b)}~the astrometric accuracy, \emph{c)}~the celestial reference frame, either HCRF\footnote{Hipparcos Celestial Reference Frame} or Gaia-CRF2 \citep{Mignard2018}, \emph{d)}~how the catalogue is tied to the International Celestial Reference System (ICRS), \emph{e)}~the coordinate epochs, \emph{f)}~the need of propagating astrometric errors when the catalogue proper motions are available and positions are given at epoch J2000.0, but errors on positions are given at mean epoch, and \emph{g)}~the known issues and caveats.
It is also important to take into account how the external catalogue properties compare to the corresponding \emph{Gaia} catalogue properties. 

Table~\ref{table:ExtProp}  lists the \emph{Gaia} DR2 and external catalogues properties relevant to the cross-match.
The effective angular resolution values reported in Table~\ref{table:ExtProp} were derived from the external catalogue reference papers or their on-line documentation. In some cases, the authors directly report the value of the effective angular resolution, in others, they list related quantities such as seeing, pixel scale, and the full width at half maximum of the point spread function (PSF FWHM), which can be used to derive
the effective angular resolution. In Subsection~\ref{subsec:resultangres} we describe the effects of the difference in effective angular resolution. Appendix~\ref{sec:app} compares the effective angular resolution values reported in Table~\ref{table:ExtProp} with the actual content of the external catalogues. In some cases, the fraction of suspected duplicates is relevant. Figures~\ref{fig:FigAngResa} and~\ref{fig:FigAngResb} are useful to understand some details of the cross-match results (see Section~\ref{sec:results}).

Figure~\ref{figure:all} shows the sky coverage and the surface density distribution for \emph{Gaia} DR2 and the external catalogues that are newly matched with  \emph{Gaia}. The corresponding figures for the external catalogues that had been matched before with  \emph{Gaia} DR1 can be found in Paper~I. The surface density is calculated by counting the number of sources in each pixel obtained using a HEALPix tessellation: for dense surveys, we adopted a resolution of $N_{\mathrm{side}}=2^{8}$ , which has 786\,432 pixels with 
a constant area of $\Omega\sim188.89~$arcmin$^{2}$, while for sparse catalogues, we adopted a resolution of $N_{\mathrm{side}}=2^{6}$ , which has 49\,152 pixels with a constant area of  $\Omega\sim0.8$~degree$^{2}$.

The external catalogue quantities used by the cross-match computations described in this study are positions, position errors, position error correlation (if available), and 
coordinate epochs. Different surveys may have a different definition of some of these quantities and/or use different units. 
The external catalogue input quantities were thus homogenised in order to simplify the cross-match calculations.

In the following we briefly describe the newly added external catalogues together with some caveats and known issues that are relevant for the cross-match computations. For catalogues that had been cross-matched with DR1, we describe some issues that were not apparent in DR1, but are relevant for the DR2 cross-match.

\subsection{Pan-STARRS1~DR1}
The Panoramic Survey Telescope and Rapid Response System (Pan-STARRS) is a system for wide-field astronomical imaging developed and 
operated by the Institute for Astronomy at the University of Hawaii. Pan-STARRS1 (PS1) is the first part of Pan-STARRS to be completed and is 
the basis for Data Release 1 (DR1). The PS1 survey used a 1.8 meter telescope and its 1.4 gigapixel camera to image the sky in five broadband 
filters (\emph{g}, \emph{r}, \emph{i}, \emph{z}, \emph{y}).
The version of the catalogue we used for cross-match computation is a filtered subsample of the $10\,723\,304\,629$ entries that are listed in the original ObjectThin table. \\
We used only ObjectThin and MeanObject\footnote{A description of the original ObjectThin and MeanObjects tables can be found at:
\url{https://outerspace.stsci.edu/display/PANSTARRS/PS1+Database+object+and+detection+tables}} 
tables to extract what we needed. This means that objects that are detected only in stack images are not included. 
The main reason for avoiding objects detected in stack images (for cross-match purposes) is that their astrometry is not as good as the mean object astrometry, as stated in the Pan-STARRS1~DR1 documentation: "The stack positions (raStack, decStack) have considerably larger systematic astrometric errors than the mean epoch positions (raMean, decMean)". The astrometry for the MeanObject positions uses \emph{Gaia}~DR1 as a reference catalogue, while the stack positions use 2MASS as a reference catalogue.

In detail, we filtered out all objects where 
\begin{itemize}
\item{nDetections = 1;}
\item{no good-quality data in Pan-STARRS, objInfoFlag 33554432 not set;}
\item{mean astrometry could not be measured, objInfoFlag 524288 set;}
\item{stack position used for mean astrometry, objInfoFlag 1048576 set;}
\item{error on all magnitudes equal to 0 or to -999;}
\item{all magnitudes set to -999;} 
\item{error on RA or DEC greater than 1 arcsec.}
\end{itemize}
The number of objects in the Pan-STARRS1~DR1 version used for cross-match is $2\,264\,263\,282$.

\subsection{APASS~DR9}
The AAVSO Photometric All Sky Survey (APASS) is obtained in five photometric bands: \emph{B}, \emph{V}, \emph{g'}, \emph{r'} and \emph{i'}, and the observed targets cover the magnitude range $10<V<17$. APASS data are obtained with dual bore-sighted 20cm telescopes, designed to obtain two bandpasses of information simultaneously,
from two sites near Weed, New Mexico, in the Northern Hemisphere and at CTIO in the Southern Hemisphere.
The APASS DR9 contains approximately 62 million stars covering about 99\%\ of the sky. The APASS project is being completed and DR9 is not a final release.
According to the APASS documentation, there are some issues in the catalogue that should be taken into account when cross-matching it:
\begin{itemize}
\item The APASS team does not provide star IDs until the final product and suggests that stars are identified by their RA and DEC.
\item There are a number of duplicate entries. These appear to be caused by the merging process, where poor astrometry in one field may cause two seed centroids to form for a single object. 
\item There are a number of entries with 0.000 errors.
\item Centroiding in crowded fields is very poor; blends cause photometric as well as astrometric errors.
\item There are saturated stars in the catalogue, and the APASS team suggests to avoid using sources brighter than $V=7$.
\end{itemize}
The issues described above are reflected in the quality of the cross-match results.

Given the lack of an identifier provided by authors, and because the Vizier TAP service\footnote{\url{http://tapvizier.u-strasbg.fr/}} is the only available resource for bulk download, we used the CDS \emph{recno} as identifier, although we are aware that the record number assigned by the VizieR team should not normally be used for identification. 

\subsection{RAVE~5}
The RAdial Velocity Experiment (RAVE) is a multi-fiber spectroscopic astronomical survey of stars in the Milky Way using the 1.2m UK Schmidt Telescope 
of the Australian Astronomical Observatory (AAO). RAVE contains multiple observations of the same source, which are identified with the same RAVEID 
and a different RAVE\_OBS\_ID. The number of entries in the catalogue is $520\,701$, while the number of distinct sources is $457\,555$. For cross-match 
calculations we used the distinct sources.

\subsection{Hipparcos2}
Hipparcos2 is a new improved reduction of the astrometric data produced by the Hipparcos mission. The astrometric accuracies are much better (up to a factor of 4) than in the original catalogue.

\subsection{Tycho-2}
The Tycho-2 catalogue is an astrometric reference catalogue containing positions, proper motions, and two-colour photometric data for the 2.5 million brightest stars in the sky. The Tycho-2 positions and magnitudes are based on precisely the same observations as the original Tycho catalogue collected by the star mapper of the ESA Hipparcos satellite, but Tycho-2 is much larger and slightly more precise, owing to a more advanced reduction technique. Components of double stars with separations down to 0.8 arcsec are included. 

There are $109\,445$ sources in Tycho-2 without an astrometric solution. These objects are indicated by {\sl pFlag}=X, where {\sl pFlag} is the mean position flag. For these objects we used the observed Tycho-2 values for coordinates, coordinate errors, and reference epoch.

There are $13\,098$ sources in Tycho-2 for which {\sl pFlag}=P. These objects are binaries (actually for 82 of them one of the two components is missing in the sample flagged with {\sl pFlag}=P) and have different source Ids, but identical astrometry since the photocentre is used for the astrometric solution, which includes proper motions.

\subsection{SDSS DR9}
A detailed description of the astrometric SDSS calibration is given in \citet{SDSSQA}, and a summary is provided in the on-line documentation\footnote{\url{http://www.sdss.org/dr12/algorithms/astrometry/}}. The \emph{r} photometric CCDs serve as the astrometric reference CCDs for the SDSS. That is, the positions for SDSS objects are based on the \emph{r} centroids and calibrations. The \emph{r} CCDs are calibrated by matching bright stars detected by SDSS with the UCAC astrometric reference catalogues.
The SDSS collaboration implemented an astrometry quality assurance (QA) system in order to identify errors in the SDSS imaging astrometry and provided a  summary file\footnote{\url{https://data.sdss.org/datamodel/files/BOSS_PHOTOOBJ/astromqa/astromQAFields.html}} containing all information about the SDSS field astrometry QA, including offsets from each of the reference catalogues. The astrometry QA summary file is available for download\footnote{\url{http://data.sdss3.org/sas/dr9/boss/photoObj/astromqa/astromQAFields.fits}}. The method we used to include the results of the above analysis in the cross-match algorithm is described in  Section~\ref{sec:results}.

\section{Cross-match output\label{sec:output}}
The cross-match output consists of two separate tables: \emph{\textup{BestNeighbour}} includes the best matches (selected as the good neighbour with the highest value of the score), while \emph{\textup{Neighbourhood}} includes all the good neighbours (selected using the second filter, see Equation~\ref{eq:secondfilter}). 
The cross-match output datamodels are described in Tables~\ref{table:BestD} and \ref{table:NeigD}. 
The content and some statistics of the \emph{\textup{BestNeighbour}} and \emph{\textup{Neighbourhood}} output tables for each external catalogue are summarised in Tables~\ref{table:Be} and \ref{table:Ne}.
\begin{table}[t]
\small
\caption{BestNeighbour output table content.}
\label{table:BestD}
\begin{tabular}{lp{0.5\linewidth}}
\hline
Field name & Short description\\
\hline
SourceId &\emph{Gaia} source identifier\\
OriginalExtSourceId & Original External Catalogue source identifier\\
AngularDistance &Haversine angular distance (arcsec)\\
NumberOfMates\tablefootmark{a}&Number of mates in \emph{Gaia} catalogue\\
NumberOfNeighbours&Number of good neighbours in the second catalogue\\
BestNeighbourMultiplicity\tablefootmark{a}&Number of neighbours with same probability as best neighbour\tablefootmark{b}\\
GaiaAstrometricParams&Number of \emph{Gaia} astrometric  \\&parameters used\\
\hline
\end{tabular}
\tablefoot{
   \tablefoottext{a}{Column available only for dense surveys. See Sections~\ref{sec:gen} and \ref{sec:details}.}
   \tablefoottext{b}{Two neighbours with the same probability are normally sources with different identifiers, but exactly the same coordinates 
   and coordinate errors. The cross-match algorithm is thus unable to distinguish them, and either can be selected as bestNeighbour.}
 }  
\end{table}
\begin{table}[t]
\small
\caption{Neighbourhood output table content.}
\label{table:NeigD}
\begin{tabular}{lp{0.5\linewidth}}
\hline
Field name & Short description\\
\hline
SourceId &\emph{Gaia} source identifier\\
OriginalExtSourceId & Original External Catalogue source identifier\\
AngularDistance &Haversine angular distance (arcsec)\\
Score & Figure of Merit\\
GaiaAstrometricParams&Number of \emph{Gaia} astrometric \\&parameters used\\
\hline
\end{tabular}
\end{table}
%
\begin{table*}
\caption{BestNeighbour statistics: Max values of relevant output fields in BestNeighbour tables. The fraction of \emph{Gaia} matched sources 
without mates and with a single neighbour as well as the number of \emph{Gaia} matched sources with no multiplicity are also listed.}
\centering
\label{table:Be}
\begin{tabular}{@{}lrcccccccc@{}}
\hline
Catalogue    & \multicolumn{1}{c}{Angular}  
                       &\multicolumn{1}{c}{Number Of}  & \multicolumn{1}{c}{\% with Single}
                       &\multicolumn{1}{c}{Number } & \multicolumn{1}{c}{\% with} 
                       & \multicolumn{1}{c}{BestNeighbour}& \multicolumn{1}{c}{Sources}\\
                      & \multicolumn{1}{c}{Distance} 
                      &\multicolumn{1}{c}{Neighbours} & \multicolumn{1}{c}{Neighbour}
                      & \multicolumn{1}{c}{Of Mates}      & \multicolumn{1}{c}{No Mates} 
                      &\multicolumn{1}{c}{Multiplicity}& \multicolumn{1}{c}{with $m>1$\tablefootmark{a}}\\ 
                      & \multicolumn{1}{c}{(arcsec)}  
                      &\multicolumn{1}{c}{ }& \multicolumn{1}{c}{}     
                      & \multicolumn{1}{c}{ } & \multicolumn{1}{c}{}& \multicolumn{1}{c}{} & \multicolumn{1}{c}{}\\ \hline
                      & \multicolumn{1}{c}{max}   
                      & \multicolumn{1}{c}{max} &\multicolumn{1}{c}{}
                      & \multicolumn{1}{c}{max} &\multicolumn{1}{c}{}
                      & \multicolumn{1}{c}{max} &\multicolumn{1}{c}{}\\  \hline
Pan-STARRS1~DR1 & 5.23                  &    6      &   98.56 &   13     &  99.42   & 1    & 0 \\                      
GSC~2.3         &   8.96                  &   16    &  96.99  &   25     &  77.65  &  16 & 120\,344\\
PPMXL          &    4.02                   &   8     &  89.35  &    14     &  87.78  &   2  & 2\\
SDSS DR9    &    52.10                &   6     &  99.49  &    80     &  99.77  &   3  & 2\\
URAT-1         &     2.12                  &   3      &  99.99  &    3      &  99.84  &  1  & 0\\
2MASS          &    5.01                  &    3      &  99.78    &     11    &  94.25            &   2  & 10 \\
allWISE         & 181.15                 &    3      &  99.99  &   24     & 98.39    &   1  & 0\\
APASS DR9 & 11.75                    &    56    &   86.53  &   59     &  58.74  &  2 & 8 \\
Hipparcos2   & 1.67                     &      2    &   99.40   & N/A     & N/A      & N/A& 0\\
Tycho-2         & 1.94                      &    3     &    99.51  & N/A      & N/A           &   N/A  & 0\\
RAVE~5        &3.21                       &    11   &     89.91  & N/A      & N/A           &    N/A  & 0\\
\hline
\end{tabular}
\tablefoot{
\tablefoottext{a}{$m$=BestNeighbour multiplicity.} 
}
\end{table*}
%
\begin{table*}
\caption{Neighbourhood statistics: Min/Max values of relevant output fields in Neighbourhood tables.}
\centering
\label{table:Ne}
\begin{tabular}{@{}lclrll@{}}
\hline
Catalogue    & \multicolumn{1}{c}{Angular Distance}  & \multicolumn{2}{c}{Score} \\
                      & \multicolumn{1}{c}{(arcsec)} &  \multicolumn{2}{c}{ } \\  \hline         
                      & \multicolumn{1}{c}{max} 
                      & \multicolumn{1}{c}{min} & \multicolumn{1}{c}{max}  \\  \hline
Pan-STARRS1~DR1    &\phantom{18} 5.25                   & 0.000000599  & 21.436935748 \\
GSC~2.3        & \phantom{18}8.96     &  0.000002516  &  21.673957559 \\
PPMXL        & \phantom{18}4.02    &  0.000004754  &  17.426431196 \\
SDSS DR9  & \phantom{1}52.42      &  0.000000076  &  17.312649378\\
URAT-1        & \phantom{18}2.12     &  0.000088545   &  18.779691209 \\  
2MASS        & \phantom{18}5.01   &  0.000013321   &  13.538666301 \\
allWISE       &  181.15                      & 0.000000676  &  15.578377337 \\ 
APASS DR9  & \phantom{18}11.75                       & 0.000002632  & 16.908079945\\
Hipparcos2   & \phantom{18}1.74                          & 0.000086919  & 20.415553214\\
Tycho-2         &\phantom{18}1.95                           & 0.000029288  & 16.896918580 \\
RAVE~5        &\phantom{18}3.44                           & 0.000001223 & 8.424990229\\
\hline
\end{tabular}
\end{table*}
%
\begin{table*}
\caption{External Catalogues cross-match results: number of objects compared with the number of matched sources, the fraction 
of distinct matched \emph{Gaia} sources and the fraction of distinct matched external catalogue sources. 
The number of sources in the neighbourhood tables is also listed.}
\centering
\label{table:Stat}
\begin{tabular}{lrrrrr}
\hline
Catalogue    & \multicolumn{1}{c}{Number of} & \multicolumn{1}{c}{Number of Best }  
                      &  \multicolumn{1}{c}{\% of \emph{Gaia}}  &  \multicolumn{1}{c}{\% of External cat}   
                      & \multicolumn{1}{c}{Number of}  \\
                      & \multicolumn{1}{c}{Sources}   & \multicolumn{1}{c}{matches\tablefootmark{a} }
                      &  \multicolumn{1}{c}{ sources matched\tablefootmark{a} }  &  \multicolumn{1}{c}{ sources matched\tablefootmark{a} }    
                      & \multicolumn{1}{c}{Neighbours}\\ \hline
Pan-STARRS1~DR1   &2\,264\,263\,282\phantom{$a$}& 810\,359\,898&80.49\tablefootmark{b} &35.68&816\,314\,072\\
GSC~2.3        &  945\,592\,683\phantom{$a$}  &   870\,899\,123   & 51.44\phantom{$a$}  & 80.96&  884\,748\,168 \\
PPMXL         &  910\,468\,688\phantom{$a$}  &   716\,220\,357   & 42.31\phantom{$a$} & 73.50&   757\,738\,601 \\
SDSS DR9  &  469\,029\,929\phantom{$a$}  &    113\,718\,207    &64.46 \tablefootmark{b}     & 24.22&    114\,011\,744  \\
URAT-1        &  228\,276\,482\phantom{$a$}  &   188\,071\,510   & 27.32\tablefootmark{b}   &82.32 &  188\,071\,646 \\  
2MASS         &  470\,992\,970\phantom{$a$}  &   450\,688\,227  & 26.62\phantom{$a$}   & 92.91 &  451\,193\,296 \\
allWISE        &  747\,634\,026\phantom{$a$}  &   300\,207\,917   & 17.73\phantom{$a$}  & 39.83 & 300\,209\,602 \\ 
APASS DR9       & 61\,176\,401\phantom{$a$}& 75\,018\,791&4.43\phantom{$a$}&90.66&81\,278\,312\\ \hline
&&&&&\\ \hline
Catalogue    & \multicolumn{1}{c}{Number of} & \multicolumn{1}{c}{Number of Best }  
                      &  \multicolumn{1}{c}{\% of \emph{Gaia}}  &  \multicolumn{1}{c}{\% of External cat}   
                      & \multicolumn{1}{c}{Number of}  \\
                      & \multicolumn{1}{c}{Sources}   & \multicolumn{1}{c}{matches\tablefootmark{c} }
                      &  \multicolumn{1}{c}{ sources matched\tablefootmark{c} }  &  \multicolumn{1}{c}{ sources matched\tablefootmark{c} }    
                      & \multicolumn{1}{c}{Neighbours}\\ \hline
Hipparcos2   & 117\,955\phantom{$a$}    & 83\,034      &0.005\phantom{$a$} &70.39&83\,283\\
Tycho-2         & 2\,539\,913\phantom{$a$}&2\,475\,900&0.15\phantom{$a$} &97.47&2\,482\,025\\
RAVE~5        & 457\,555\tablefootmark{d}&450\,587&0.027\phantom{$a$} &98.48&474\,824\\
\hline
\end{tabular}
\tablefoot{
   \tablefoottext{a}{Column "Number of Best matches" includes the \emph{mates}. This column and column "\% of \emph{Gaia}  sources matched" indicate distinct matched \emph{Gaia} sources. Column "\% of External cat sources matched" indicates the fraction of distinct external catalogue sources that were matched.}
   \tablefoottext{b}{The percentage of matched \emph{Gaia} sources in this case takes into account the limited sky coverage of the external catalogue (see Figure~\ref{figure:all}).}
  \tablefoottext{c}{Column "Number of Best matches" does not include the \emph{mates}, since for sparse catalogues a one-to-one best match is forced.
  Column "\% of \emph{Gaia}  sources matched" indicates distinct \emph{Gaia} sources. Column "\% of External cat sources matched" indicates the fraction of distinct external catalogue sources that were matched.}
   \tablefoottext{d}{RAVE contains multiple observation of the same source, which are identified with the same RAVEID and a different RAVE\_OBS\_ID. The number reported in this table is the number of RAVE distinct sources.}
   }
\end{table*}
%
\section{Results\label{sec:results}}
The cross-match results are part of the official \emph{Gaia} DR2 release and are available at the ESA Gaia Archive\footnote{\url{https://gea.esac.esa.int/archive/}} and at Partner Data Centres Archives\footnote{Space Science Data Center - ASI (\url{http://gaiaportal.ssdc.asi.it/}), Leibniz Institut f\"ur Astrophysik Potsdam - AIP (\url{https://gaia.aip.de/}),  Astronomisches Rechen-Institut (\url{http://gaia.ari.uni-heidelberg.de/})}. 
The cross-match results are described in Table~\ref{table:Stat} and in Figs.~\ref{fig:FigResulta}- \ref{fig:FigAngDista}. 

Given the size of the catalogues involved in this cross-match study, the analysis of the results can be performed only on general grounds, certainly not on an object-by-object basis. The aim of the following analysis is thus to give users information on the global characteristics of the cross-match results for a given catalogue, that is, sky and magnitude distributions of matched sources, distribution of angular distance of matched pairs, and fraction of matched sources.

In particular, the surface density maps displayed in the left column of Figure~\ref{fig:FigResulta} show the fraction of matched \emph{Gaia} sources, while the maps in the
right column show the fraction of matched external catalogue sources. These maps, combined with the corresponding maps available in Figure~\ref{figure:all} or in Paper~I, allow the spatial analysis of the cross-match results. In the case of GSC~2.3 and PPMXL, the cross-match with the duplicated sources located at the plate borders results in an over-density of matched \emph{Gaia} sources that is clearly visible in Figure~\ref{fig:FigResulta} as a square pattern.

The histograms in Figure~\ref{fig:FigHist} show the magnitude distribution of the matched external catalogue sources compared with the distribution of the full catalogue. Figure~\ref{fig:FigHist} can thus be used to assess the fraction of matched and missed external catalogue sources as a function of magnitude.
                               
The angular distance distributions shown of cross-matched pairs in Figure~\ref{fig:FigAngDista}  can be used to evaluate the global agreement of the external catalogue astrometry with \emph{Gaia}. They can be used to retrieve information about the angular distance at which the bulk of the matched pairs are found (blue histograms) and about the angular distance within which all the matched sources are found (cumulative red curves). In addition, they also show no indication of the Poisson tail that is always present in cone search results: one of the advantages of a cross-match over a cone search is indeed that the search radius is defined on a pair-by-pair basis and is not fixed for all pairs. 
For example, even if in Figs.~\ref{fig:FigResulta} and  \ref{fig:FigHist} APASS DR9 and 2MASS cross-match results show similar behaviours, it is instead clear from
Figure~\ref{fig:FigAngDista} that the 2MASS  positions are in much better agreement with \emph{Gaia} than the APASS DR9 positions.
It is important to note that the histograms in Figure~\ref{fig:FigAngDista} are not a direct indication of the astromentric quality of the external catalogues. The main reason is that the histograms show only the matched sources, while astrometric issues in a catalogue often prevent the match of a fraction of the potential counterparts, leaving only the sources with good astrometry. In this study, by "good astrometry" we mean not only accurate positions, but also a careful evaluation of the position errors and the inclusion of systematics in position errors. The cross-match algorithms that require counterparts to be compatible within position errors easily highlight when position errors are underestimated.
In the following we illustrate and discuss some specific features and characteristics of the cross-match results.

 \subsection{Effect of effective angular resolution differences on XM results\label{subsec:resultangres}}
The comparison between the effective angular resolution of  \emph{Gaia} and of the external catalogues is very important for the cross-match. The higher  \emph{Gaia} angular resolution (which will 
improve with the forthcoming releases) implies that  \emph{Gaia} will frequently resolve sources that are unresolved in the external catalogue. The larger the difference in effective
resolution between  \emph{Gaia} and the external catalogue, the more common the resolved objects.
For this reason, since  \emph{Gaia} DR1 we chose a many-to-one algorithm for dense surveys and defined the mates, which are two or more \emph{Gaia} sources with the same best neighbour in the external catalogue. The external catalogue sources that are the counterpart of two or more \emph{Gaia}  sources are thus very likely sources that are resolved in \emph{Gaia}.  The chances that mates correspond to a resolved object are obviously higher when all mates have proper motions available, and thus their positions are reliably propagated to the external catalogue epoch. A more subtle effect arises when the photocentre of the unresolved external catalogue source is too far from the corresponding two or more  \emph{Gaia} counterparts to allow a match within position errors. For these cases, a complex dedicated treatment is required.
While  the released \emph{Gaia} DR2 data, and in particular the availability of accurate five-parameter astrometric solutions, allow addressing the angular resolution difference effects on cross-match, the solution is not trivial and requires carefully planned tests and  a thorough analysis. A detailed treatment of this effect will be included in the cross-match of \emph{Gaia} DR3 and subsequent releases.

\subsection{Hipparcos2\label{subsec:hipissue}}
While we expect to find \emph{Gaia} counterparts for most of the Hipparcos2 sources, with the exception of the brightest ones, the cross-match results include only $\text{about two-thirds}$ of them. This means that according to the adopted cross-match algorithm, only $\text{about two-thirds}$ of the Hipparcos2 objects have a \emph{Gaia} counterpart that is compatible within the position errors (i.e. have at least one good neighbour). 
Hence the Hipparcos2 cross-match results clearly show an issue that needs to be investigated. 

Around each Hipparcos2 object, we calculated a cone search with a fixed radius of 1~arcsec, which propagates the \emph{Gaia} positions to Hipparcos2 epoch exactly in the same way as in the cross-match algorithm described in this paper. Then we selected the nearest neighbour. The cone search is thus consistent with the cross-match and allows us to make a direct comparison of the angular distance distributions obtained with the two methods that were used to determine possible counterparts. 
We defined two subsamples of Hipparcos2 sources. The first includes the cross-matched sources, and the second the additional associations that were found using the cone search. 
We then tried to identify a characteristic (either in Hipparcos2 or in \emph{Gaia}) that could be used to separate the two samples and thus to understand the nature of the considered issue.
The two samples are indistinguishable in terms of size of astrometric errors (see \citet{Lindegren2018}, Appendix A, and \citet{Arenou2018} Subsection~4.6, for a detailed discussion of \emph{Gaia} astrometric errors), magnitude or colour distribution, sky distribution, and many other quantities listed in the Hipparcos2 and \emph{Gaia} catalogues. The only parameters that seem on average to allow separating the two samples are parameters related to the \emph{Gaia} astrometric solution quality, for example the \emph{astrometricGofAl} (goodness-of-fit statistics of the astrometric solution for the source in the along-scan direction). 

The top panel of Figure~\ref{Fig:HipIssue} shows the angular distance distribution of the cross-matched sample (red histogram) and the sample of additional sources added with the cone search (blue histogram). 
The panel clearly shows that the blue sample Hipparcos2 sources are found at larger angular distances from their Gaia counterparts than the red ones. The blue sample associations are found at an average angular distance of 75.4~mas, while the red sample sources are found at an average angular distance of 13.8~mas.
The middle panel of Figure~\ref{Fig:HipIssue} shows the \emph{astrometricGofAl} distribution of the cross-matched sample and the sample of associations added with the cone search, but only for \emph{Gaia} sources with a five-parameter astrometric solution.  
The bottom panel of Figure~\ref{Fig:HipIssue} shows the sky distribution of the \emph{astrometricGofAl} averaged over healpix obtained with an HEALPix tessellation with resolution $N_{\mathrm{side}}=2^{8}$ for the  \emph{Gaia} catalogue sources with a five-parameter astrometric solution.
The \emph{astrometricGofAl} sky distribution allows a comparison between the values of the \emph{astrometricGofAl} for the two samples with values of the \emph{Gaia}~DR2 catalogue. 

The adopted cross-match algorithm does not account for the effects that arise because Hipparcos2 and \emph{Gaia} DR2 have different reference frames, HCRF and Gaia-CRF2, respectively.
According to  Subsection~5.1 of \citet{Lindegren2018}, the  global alignment of  Gaia-CRF2 evaluated by the frame orientation parameters [$\epsilon_{X},\epsilon_{Y},\epsilon_{Z}$] at J2015.5 is constrained within $\pm$0.02  mas  per  axis for faint sources, and there is no indication of a misalignment larger than  $\pm$0.3 mas per axis at the bright end.
The Hipparcos2 misalignment at epoch J1991.25 is $\pm$0.6 mas per axis. 
Concerning  the  spin  of  the  reference  frame  relative  to the quasars,  \citet{Lindegren2018} confirmed that the faint reference frame of Gaia DR2 is globally non-rotating to within $\pm$0.02 mas/yr in all three axes. However, using a subsample of the Hipparcos2 sources present in TGAS (Tycho-Gaia Astrometric Solution, the subsample of \emph{Gaia} DR1 sources with a five-parameter astrometric solution), the authors suggested that  the  bright  (G$\lesssim$12)  reference  frame  of Gaia DR2  has  a  significant  ($\sim$0.15  mas/yr) spin  relative  to  the  fainter  quasars. According to them, the most reasonable explanation is systematics in the Gaia DR2 proper motions of the bright sources. 
The effects of either the combination of HCRF and Gaia-CRF2 misalignments or the inertial spin of the Gaia DR2 proper motion system 
are too small  when compared to the bulk of angular distances between Hipparcos2 sources associated with the cone search and their Gaia counterparts. These effects therefore cannot account for the bulk of the missing Hipparcos2 matches.

The cross-match is particularly critical when two catalogues with such small positional uncertainties are combined. 
The explanation for the one-third of Hipparcos2 sources without a \emph{Gaia} counterpart 
compatible within position errors seems to reside in non-optimal astrometric 
solutions for part of the Hipparcos2 sources, as a result of astrometric perturbations that are probably caused by multiplicity, variability, and/or peculiarities.

Since we do have the a priori knowledge that we should match almost all the Hipparcos2 sources, we decided to add the result of the 1~arcsec cone search described above to this paper and make it available to users for download\footnote{\url{https://www.cosmos.esa.int/web/gaia/dr2-known-issues}}. The table contains three  columns: the \emph{Gaia} and Hipparcos2 identifiers, and the angular distance (in arcsec) for each nearest associated source. Table~\ref{table:cone} contains the first ten entries of the cone search results.

\begin{table}
\caption{Sample of the cone search results described in Subsection~\ref{subsec:hipissue}}
\centering
\small
\label{table:cone}
\begin{tabular}{@{}rrl@{}}
\hline
\multicolumn{1}{c}{Gaia} & \multicolumn{1}{c}{Hipparcos2 }  & \multicolumn{1}{c}{Angular Distance} \\
\multicolumn{1}{c}{SourceId} &\multicolumn{1}{c}{Identifier} & \multicolumn{1}{c}{(arcsec)}  \\  \hline   
5188150893900488576&48752 &0.002032030044478262 \\
5764614467999340032&71348 &0.015010173408366224 \\
5188178214189131008&42708 &0.0033109008787438705\\
5764662880870489728&78866 &0.006914112668101691 \\
4611734916632361600&22645 &0.002173230360375837 \\
5188197627441445632&54065 &0.24059879382091026  \\
4611782058193541248&3560  &0.0056529407401837315\\
6341351575677860992&90987 &0.11695414581743536  \\
5188247891443554688&40104 &0.2276057244922324   \\
6341181494973204096&104382&0.021732003615886262 \\                            
\hline
\end{tabular}
\end{table}

\subsection{Tycho-2}
As described in Section~\ref{sec:extcat}, for the cross-match computations we preferentially used the Tycho-2 set of coordinates propagated to epoch J2000.0.  For a fraction of binary sources resolved by Tycho-2, however, the photocentre of the binary was used to obtain the astrometric solution (and thus the binary components have the same coordinates). Since Tycho-2 binaries have separations larger than $\sim$0.8 arcsec, they should also be resolved by \emph{Gaia}, even if both components are not always present in the \emph{Gaia}~DR2 catalogue. In these cases, which involve $13\,098$ sources, the cross-match results are very poor, and we matched only  3744 sources. In these cases, both components are included in the cross-match output and will obviously both have the same \emph{Gaia} counterpart(s). This problem will be addressed for DR3, when we will use the Tycho-2 observed positions, which are given separately for different components and allow a greatly improved number of binary matches.

\subsection{GSC~2.3 and PPMXL}

The GSC~2.3 and PPMXL catalogues can be considered similar since they are both based on the same photographic plates, but  PPMXL has a composite nature (see the Introduction of \citealt{PPMXL}). The PPMXL coordinates that are available in the original catalogue were propagated to J2000.0. For cross-match purposes, we computed the position errors at J2000.0 using the position errors at mean epoch available in PPMXL. Nevertheless, the GSC~2.3 position errors are typically four times larger than the propagated PPMXL errors. 
According to \citealt{GSC2.3}, GSC~2.3 position errors should be considered conservative estimates of the uncertainties. The epoch difference between \emph{Gaia} and GSC~2.3 is $\sim$25 years on average, and it is 15.5 years for PPMXL.

Given the above, the cross-match results are quite different for the two catalogues.
First of all, $\sim$81\% of GSC~2.3 sources and only 73.5\% of PPMXL sources have a \emph{Gaia} counterpart.
Figure~\ref{fig:FigAngDista} shows that when \emph{Gaia} counterparts are found, PPMXL sources are closer than GSC~2.3 sources, 
but this does not mean that the PPMXL astrometry is better than GSC~2.3 astrometry. 
The longer the time interval for which a given \emph{Gaia} source is propagated, the larger is the possible misplacement, due to proper motion uncertainties. This explains in part why GSC~2.3 counterparts are found at larger distances than PPMXL counterparts.
On the other hand, since PPMXL positions are propagated to J2000.0 using PPMXL proper motions, when they are not accurate, the \emph{Gaia} counterparts are less easy to find.
The PPMXL small position errors also contribute to the counterpart matching failures, while the larger GSC2.3 position errors allow us to find counterparts at larger distances.
The net effect is that \emph{Gaia} counterparts are fewer but closer in PPMXL and more numerous but at larger distances in GSC2.3. 

The similarity of the two catalogues instead accounts for the similar issue with duplicates at plate edges and for the similar secondary feature shown in Figure~\ref{fig:FigAngDista}, roughly between 0.8 and 1.8 arcsec, which is due to the presence of mates in cases when one of the two different \emph{Gaia} sources that share the same best neighbour in the external catalogue is much closer than the other.
The duplicate issue is more evident for PPMXL, while the described secondary feature is more distinguishable for GSC~2.3, see Figure~\ref{figure:all}.
\subsection{2MASS and allWISE}
Both 2MASS and allWISE have good astrometry and show no strong indications of an issue with duplicated sources.
In particular, 2MASS does not show signatures of astrometric problems or position error underestimation in the maps included in Figure~\ref{fig:FigResulta}. 
As detailed in Table~\ref{table:Be}, the 5.75\% of the \emph{Gaia} sources that match a 2MASS source have a mate (i.e. are resolved in \emph{Gaia}), this means that the 2MASS 
cross-match probably already includes most of the \emph{Gaia} resolved objects and will not benefit much from the more detailed treatment foreseen for DR3 (see Subsection~\ref{subsec:resultangres}).   

Of the external catalogues included in this study, allWISE is the farthest in the infrared and has the lowest angular resolution. Another characteristic of allWISE is that the Galaxy is less prominent in its surface density distribution (see Figure~2 in Paper~I), meaning that the surface density distribution variations are lower. These three characteristics explain why the fraction of matched Gaia sources and the fraction of matched allWISE sources are both small. Given its low angular resolution and the relatively small position errors, allWISE will probably appreciably benefit from the more detailed treatment of resolved \emph{Gaia} sources, but it will not dramatically increase the number of matches.
\subsection{SDSS DR9 and Pan-STARRS1~DR1}
SDSS~DR9 and Pan-STARRS1~DR1 are similar as both are deeper than \emph{Gaia}, are observed in the same photometric system, and have comparable angular resolutions ($\sim$0.7 and $\sim$1.1 arcsec respectively).
However, the Pan-STARRS1~DR1 position errors are definitely smaller than those of  SDSS~DR9. In the case of Pan-STARRS1~DR1, 90\% of the objects have position errors smaller than $\sim$120  mas, while in the case of SDSS~DR9, the position errors of the 90\% of sources are smaller than $\sim$250 mas.

After the first attempts to cross-match \emph{Gaia}~DR2 with SDSS~DR9, we realised that we were matching too few objects compared to what we obtained for DR1.
We thus decided to use the astrometry QA summary file described in Section~\ref{sec:extcat}, and in particular, the listed astrometric differences in RA and DEC with respect to UCAC-3, which are average differences within a given field. 
We computed the standard deviation considering all fields obtaining $\sim$50 mas in RA and $\sim$70 mas in DEC, and we thus applied a systematic that is common to all SDSS sources, ameliorating the general cross-match. For \emph{Gaia}~DR3, we will apply for each SDSS source the systematic of the corresponding field, and we will use SDSS~DR13, which has a new improved photometric calibration with respect to DR9.

Figure~\ref{fig:FigAngDista} shows that the angular distance distribution of matched sources is narrower for Pan-STARRS1~DR1 than for SDSS~DR9 and that the peak is closer to zero in the case of Pan-STARRS1~DR1.
However, the maps in panels a) and g) of Figure~\ref{fig:FigResulta} clearly show a different pattern for SDSS~DR9 and Pan-STARRS1~DR1.
In the case of SDSS~DR9, most \emph{Gaia} sources are matched as expected, given SDSS~DR9 has an higher photometric depth. In the case of Pan-STARRS1~DR1, in contrast, a fraction ($\sim$30\%) of \emph{Gaia} sources at high Galactic latitudes are not matched, even though these sources were observed by Pan-STARRS1. An in-depth analysis of the characteristics of matched and not matched Pan-STARRS1~DR1 sources shows that the 
cause might be an issue with the Pan-STARRS1~DR1 astrometric calibration at high Galactic latitudes, which is highlighted by the small position errors. The Pan-STARRS1~DR1 astrometric calibration is described in \citet{panstarrs1e}.
It should also be noted that the position error broadening method used in the cross-match algorithm described in this paper (see Subsection~\ref{sub:epochdiff}) implies that when the unknown proper motion of a given
\emph{Gaia} source is small, it is somewhat easier to find a match for that source compared with the \emph{Gaia} sources with a full five-parameter astrometric solution.  
As a direct consequence, the cross-match result for Pan-STARRS1~DR1 includes the fainter \emph{Gaia} sources, which constitute the bulk of \emph{Gaia} sources without proper motions.
For both SDSS~DR9 and Pan-STARRS1~DR1, the histograms included in Figure~\ref{fig:FigHist} show that the cross-matched sample correctly does not include the faint sources, which are not observed by \emph{Gaia}.
\subsection{URAT~1}
URAT~1 is shallower than \emph{Gaia}, has a larger effective angular resolution, and the position errors for most sources are smaller than 25 mas.
The cross-match results summarised in Tables~\ref{table:Be} and \ref{table:Stat} indicate that $\sim$82\% of URAT~1 sources have a counterpart in \emph{Gaia} and that in most cases, a single \emph{Gaia} object is matched to a given URAT~1 source (i.e. most matched \emph{Gaia} sources do not have mates).
The angular distance distribution in Figure~\ref{fig:FigAngDista} shows that the peak is very close to zero ($\sim$0.027 arcsec), but panel j) of Figure~\ref{fig:FigResulta} shows that regardless of the shallowness, not all URAT~1 sources have a \emph{Gaia} counterpart. This can be due to either small astrometric issues, or most probably, to the fact that position errors are underestimated to some degree. The 
2MASS survey is also shallower than \emph{Gaia,} and the map that shows the fraction of its sources that matched \emph{Gaia} (panel l) of Figure~\ref{fig:FigResulta}) can be   
compared with the corresponding URAT~1 map. It is clear that the astrometry for 2MASS agrees better with \emph{Gaia}, as confirmed by the total of 2MASS matched sources  ($\sim$93\%).

\subsection{APASS~DR9}
The analysis of APASS~DR9 cross-match results clearly shows that this survey is affected by various issues, such as the anomalously low steepness of the the cumulative percentage angular distance of matched pairs shown in Figure~\ref{fig:FigAngDista}.
Even though the algorithm found a match for more than 90\%\ of the APASS sources with \emph{Gaia}, a significant fraction of them consists of duplicated sources. 
This assertion is supported by the angular distance distribution of the nearest neighbours shown in Figure~\ref{fig:FigAngResa}, where it is evident that a large portion of APASS sources ($\sim$8.7 million) have at least one neighbour located at a smaller spatial scale than the angular resolution of the survey ($\sim$5~arcsec). In contrast to GSC~2.3, PPMXL, and Pan-STARRS1~DR1, in the case of APASS the presence of duplicates is a more 
general problem and they are present not only at tile edges. Hence, it is important to recall that DR9 is not the final release of the APASS project, and the cross-match results should be used with particular caution.

\subsection{RAVE~5}
When we analysed the angular resolution of the external catalogues, we found 5633 pairs of sources and 13 triplets of sources in RAVE~5, which, while having different RAVEIDs, seem to be the same sources and are found at distances closer than 3.0 arcsec from each other (and which can be easily found using a cone search). Since RAVE~5 is a sparse catalogue and the cross-match algorithm we use for sparse catalogues forces a one-to-one match, only one of the sources belonging to a given pair or triplet is matched with a \emph{Gaia} source.

   \begin{figure*}
   \centering
   \includegraphics{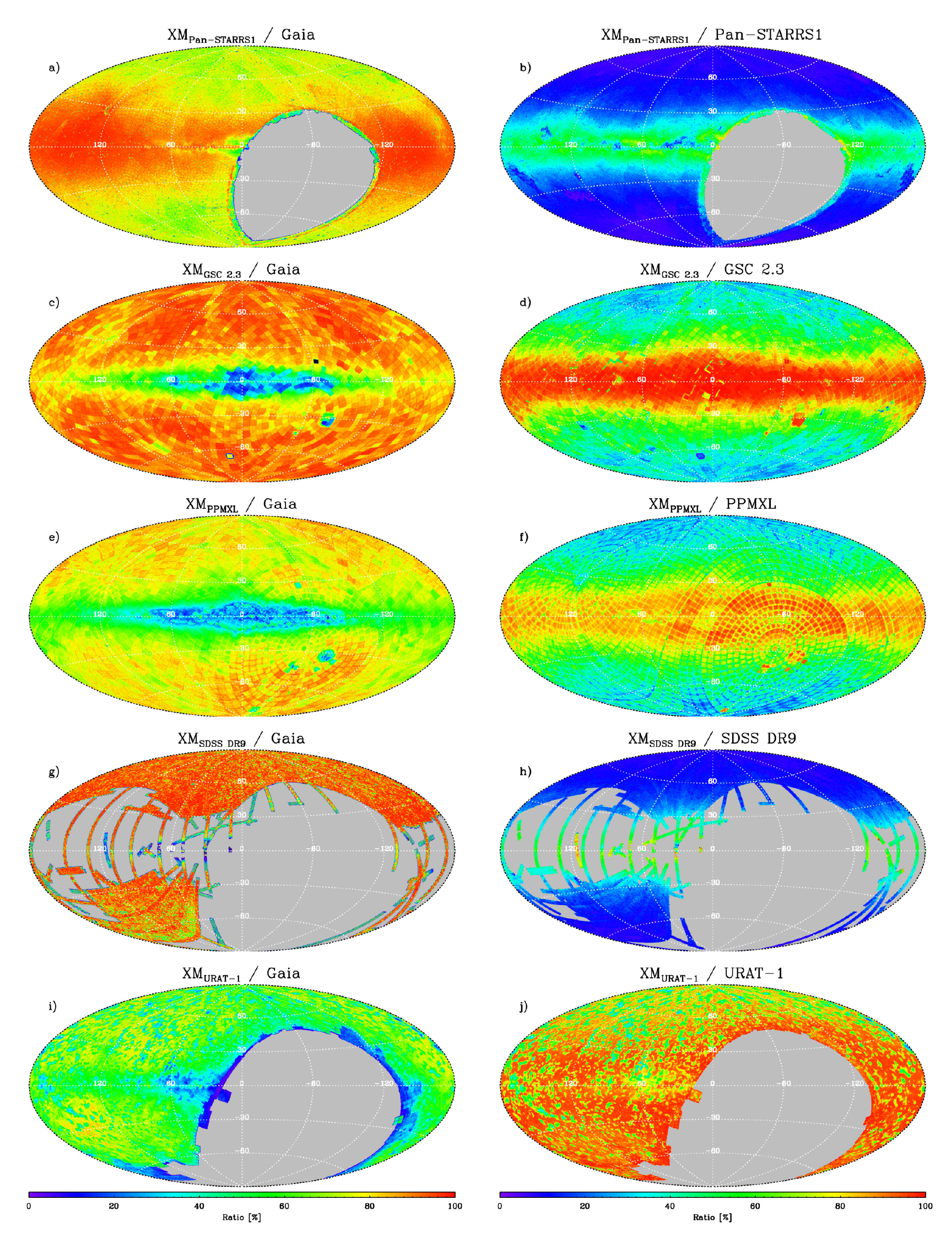}
   \caption{Surface density map for matched sources obtained using a HEALPix tessellation with resolution $N_{\mathrm{side}}=2^{8}$ for dense surveys. Left column figures show the fraction of \emph{Gaia} sources that match with an external catalogue, while the right
   column figures show the fraction of distinct external catalogue sources that match with \emph{Gaia}. }
              \label{fig:FigResulta}
  \end{figure*}
  %
   \begin{figure*}
    \setcounter{figure}{3}
   \centering
   \includegraphics{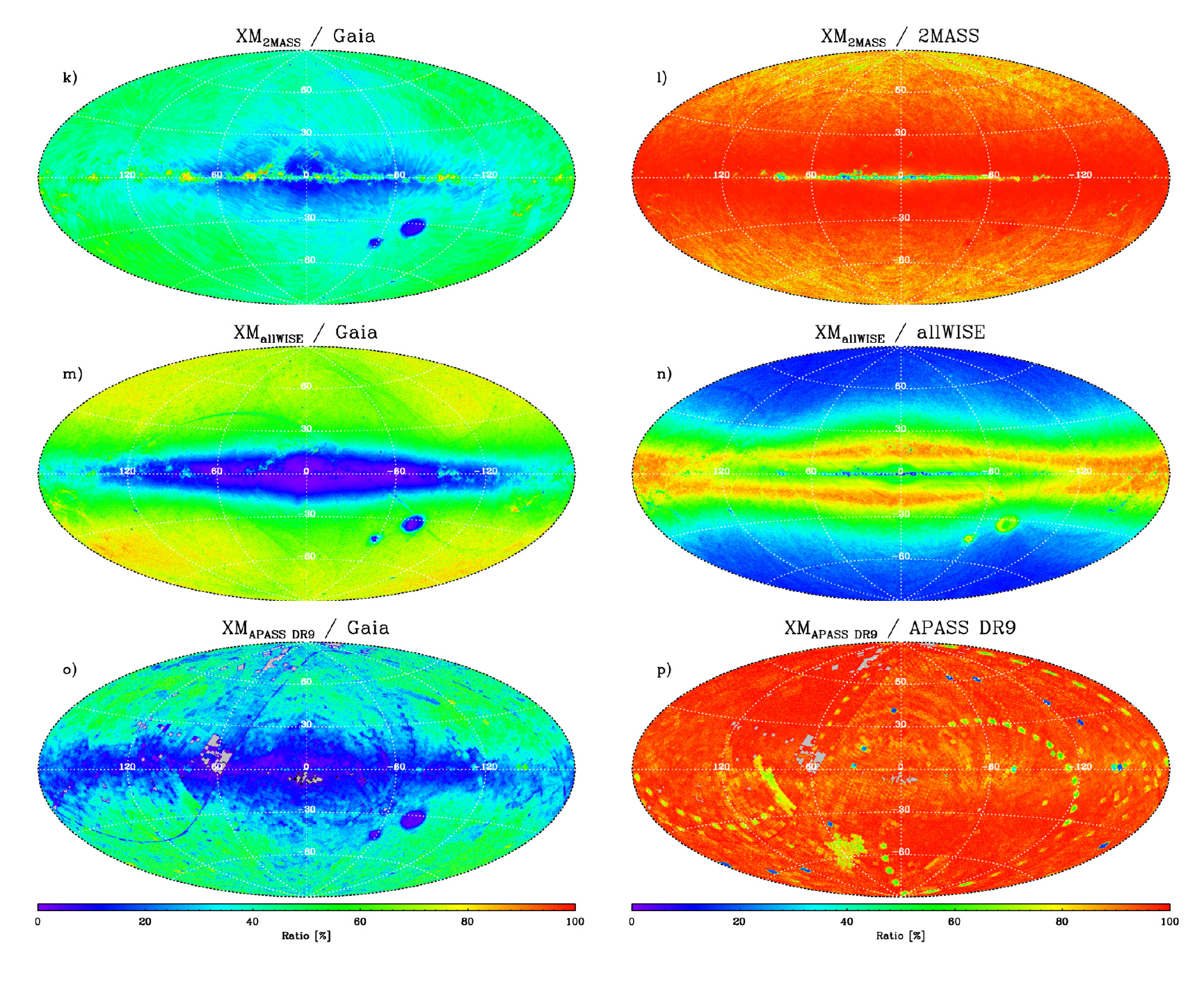}
   \caption{continued.}
              \label{fig:FigResultb}
    \end{figure*}
    %
    
   \begin{figure*}
   \centering
   \includegraphics[width=0.96\linewidth]{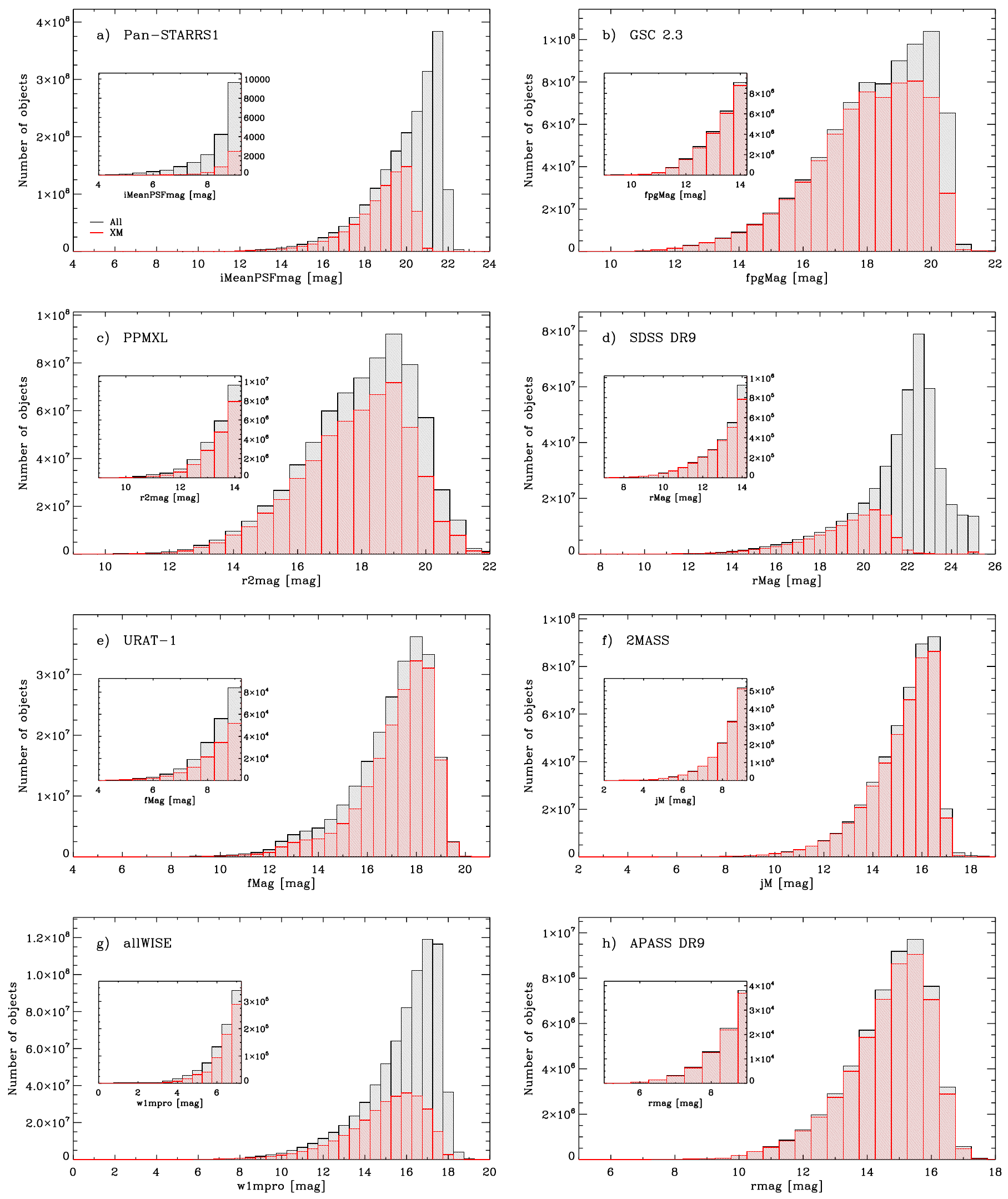}
   \caption{Magnitude distribution in the most populated band for the sources in the external catalogues. 
   In grey we plot the catalogue distribution and in red the matched source distribution.}
              \label{fig:FigHist}%
    \end{figure*}
   %

   \begin{figure*}
   \centering
  
   \includegraphics[width=0.96\linewidth]{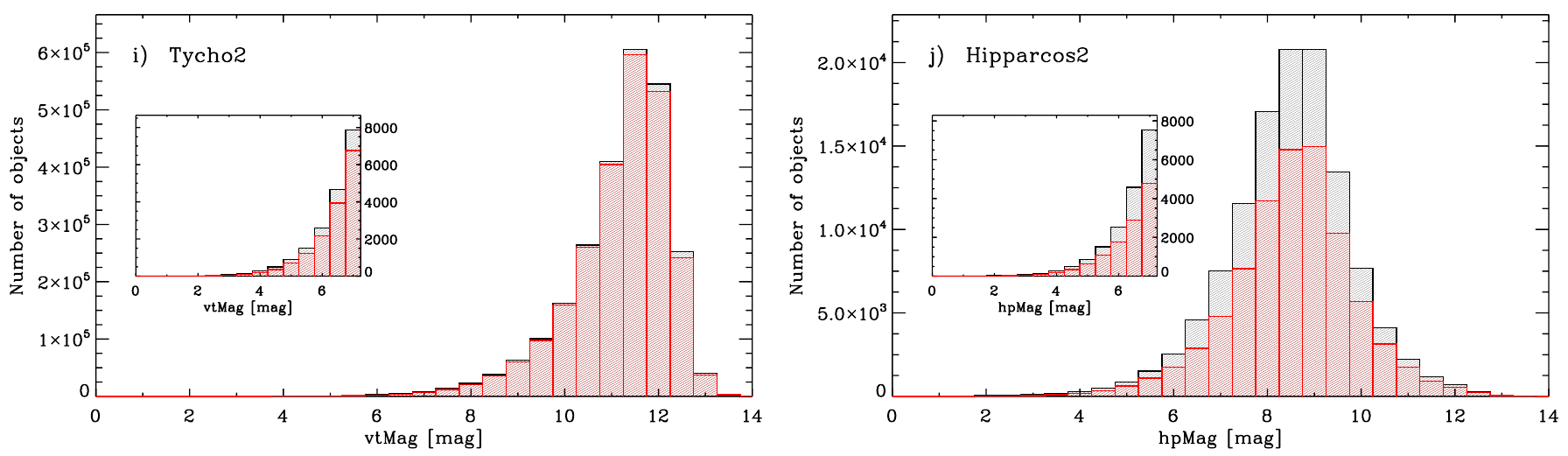}
    \setcounter{figure}{4}
   \caption{continued.}
              \label{fig:FigHist2}%
    \end{figure*}
%

   \begin{figure*}
   \centering
   \includegraphics[width=0.96\linewidth]{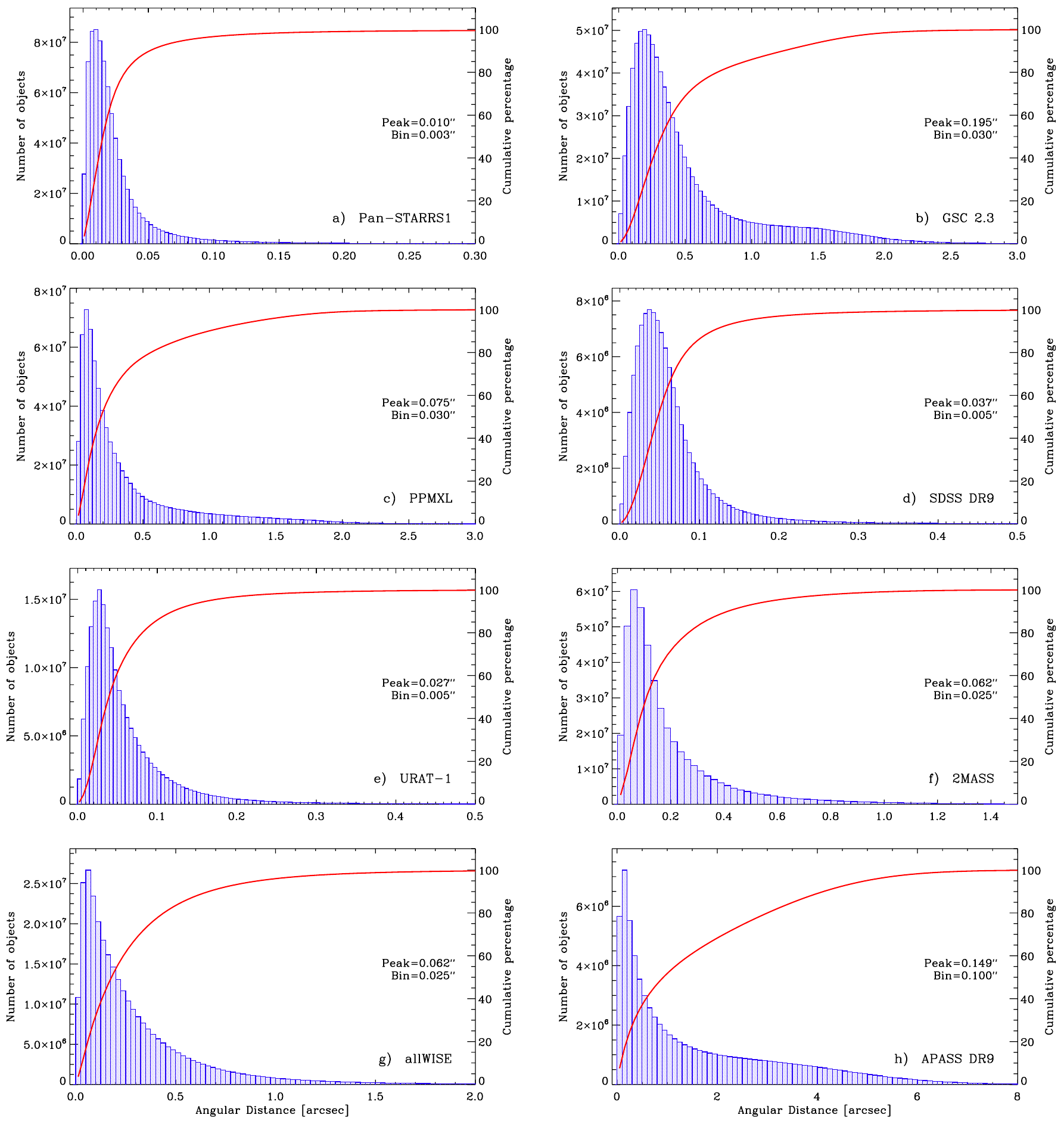}
   \caption{Angular distance distribution of cross-matched pairs included in the BestNeighbour tables. The red curves represent the cumulative distribution of the angular distance.}
              \label{fig:FigAngDista}
  \end{figure*}
  %
  
   \begin{figure}
    \setcounter{figure}{5}
   \centering
   \includegraphics[width=0.96\linewidth]{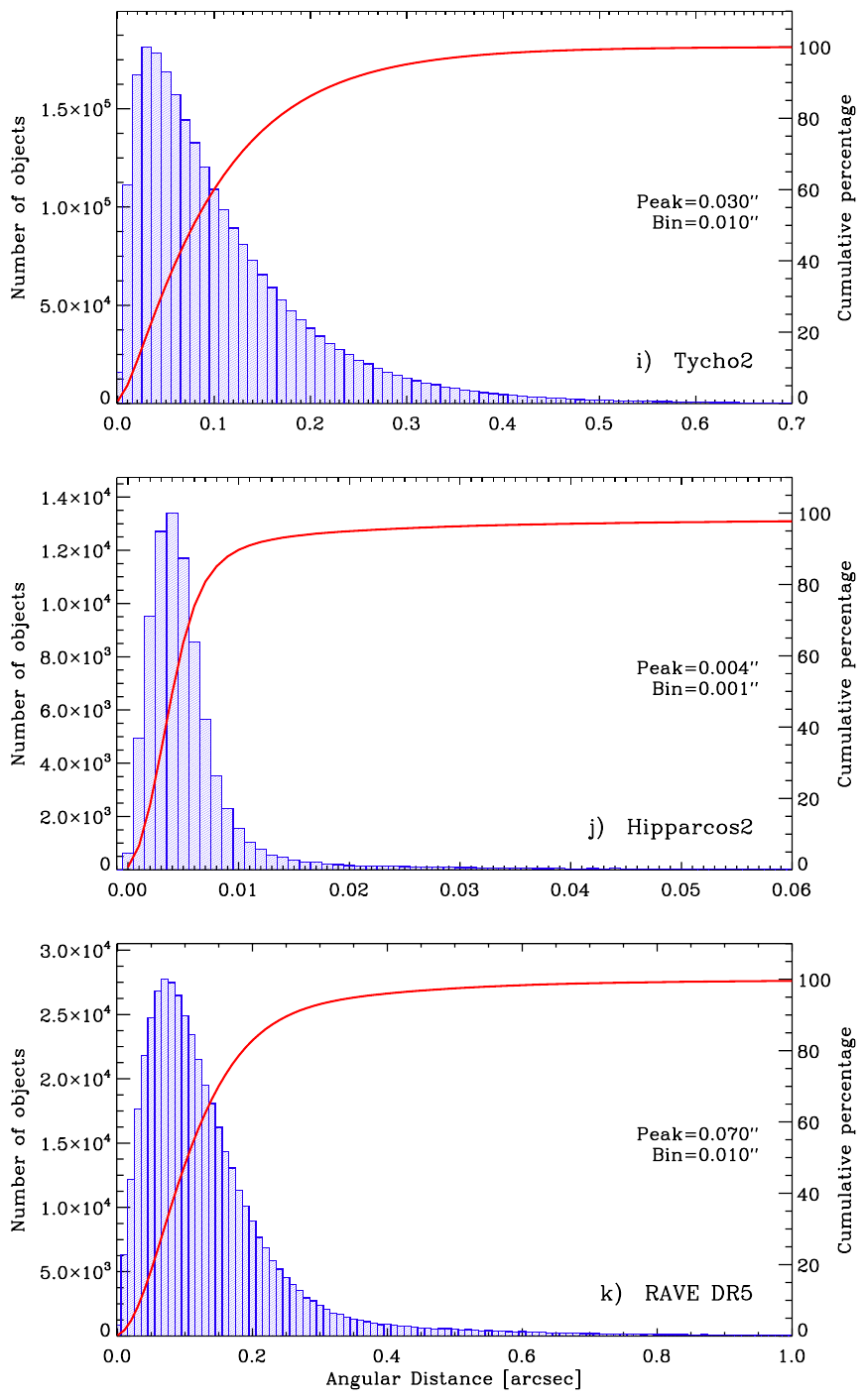}
   \caption{continued.}
              \label{fig:FigAngDistb}
    \end{figure}

\section{Conclusions\label{sec:final}}

  \begin{figure}
   \centering
   \includegraphics[width=1.0\linewidth]{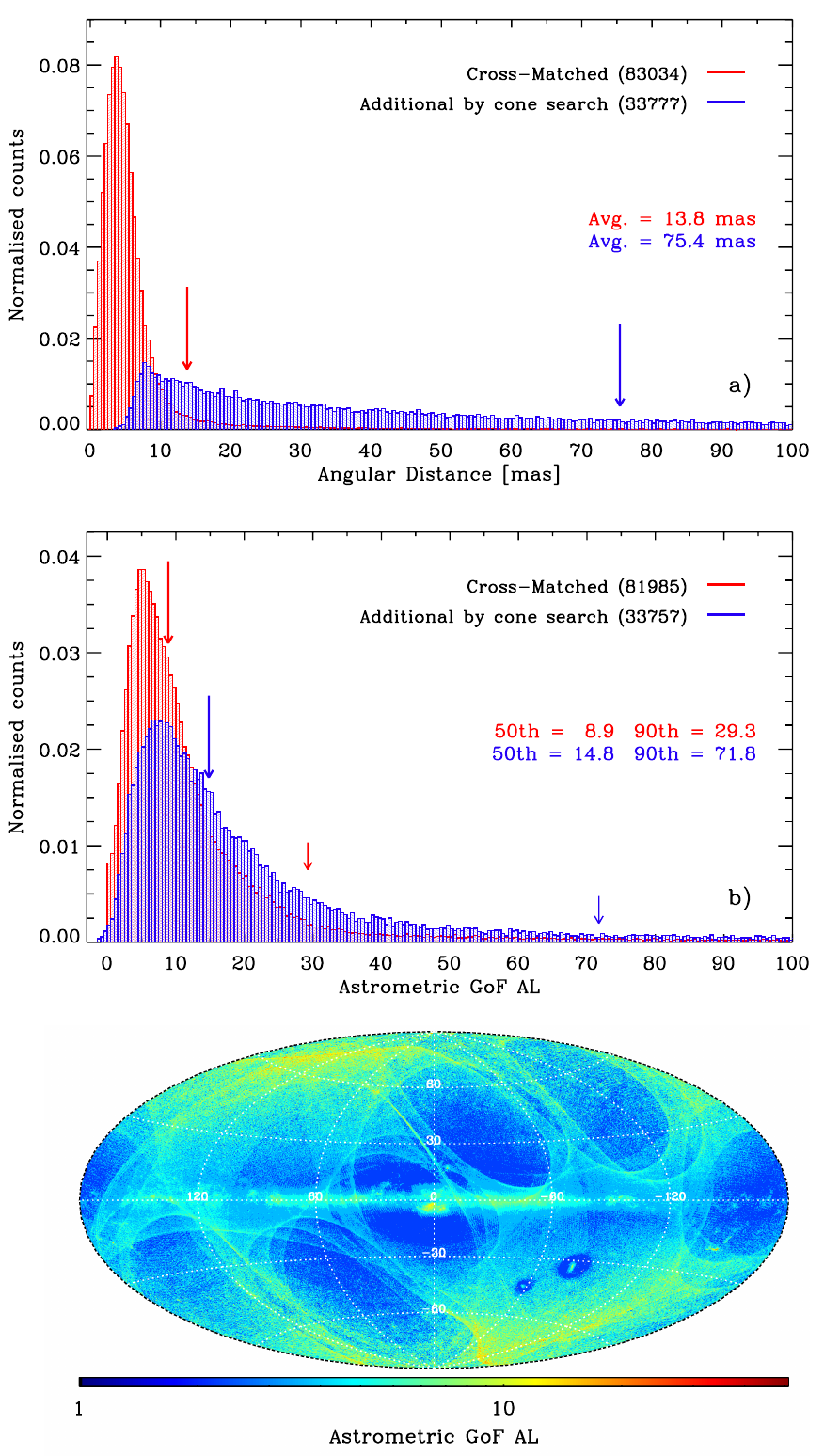}
   \caption{Issues encountered when cross-matching Hipparcos2 with \emph{Gaia}. For a detailed description and explanation of the results shown in this figure, we refer to the main text (Subsection~\ref{subsec:hipissue}).} 
              \label{Fig:HipIssue}
    \end{figure}
We presented the algorithms we developed for the official cross-match of the high-accuracy \emph{Gaia} DR2 astrometric data with eight large dense surveys and three sparse catalogues. The defined algorithms are positional and are able to fully exploit the enormous number of \emph{Gaia} sources with accurate proper motions and parallax measurements using the full five-parameters astrometric covariance matrix on an object-by-object basis. In addition, we included an improved definition of the surface density of observed objects for each catalogue, which allows a better evaluation of the local environment.

 The external catalogues and cross-match results were also described. In particular, we analysed the global behaviour of the cross-match results by evaluating their sky distribution, statistical indicators, magnitude, and angular distance distributions. More importantly, we tried to supply scientists, both in the output tables and in the analysis performed in this paper, with all the means to verify the quality of the cross-match results and to understand whether this cross-match is appropriate for their scientific needs.

The excellent data provided by the \emph{Gaia}~DR2, and in particular the proper motions, substantially improve the quality of \emph{Gaia} counterparts that are found in external catalogues. The high accuracy of the current \emph{Gaia} data gives a strong drive and powerful tools for understanding and quantifying known complex issues (such as resolution effects, and the presence of astrometric binaries and of duplicated sources) that influence the cross-match results and require non-trivial solutions. The issues will be tackled in the forthcoming \emph{Gaia} data releases. 
\begin{acknowledgements}
It is a pleasure to acknowledge the anonymous referee for their contribution to  the improvement  of the content and the readability of our manuscript.
We would like to acknowledge the financial support of INAF (Istituto Nazionale di
Astrofisica), Osservatorio Astronomico di Roma,  ASI (Agenzia
Spaziale Italiana) under contract to INAF: ASI 2014-049-R.0 dedicated to SSDC.

This work has made use of data from the European Space Agency (ESA)
mission {\it Gaia} (\url{https://www.cosmos.esa.int/gaia}), processed by
the {\it Gaia} Data Processing and Analysis Consortium (DPAC,
\url{https://www.cosmos.esa.int/web/gaia/dpac/consortium}). Funding
for the DPAC has been provided by national institutions, in particular
the institutions participating in the {\it Gaia} Multilateral Agreement.
 We would like to thank  G.~Fanari, D.~Bastieri, B.~Goldman, R.A.~Power, C.~Babusiaux, F.~Arenou, C.~Fabricius, L.~Inno, C.~Bailer-Jones, and T.~Zwitter for very useful discussions, suggestions, help, and support. 
\end{acknowledgements}


\bibliography{marrese}
\clearpage
\begin{appendix}

\section{Effective angular resolution\label{sec:app}}
The effective angular resolving power (or resolution) that results from combining a telescope and its detector is the smallest angle between close objects that can be seen to be separate. 
The effective angular resolution can be ill defined in astronomy for various reasons: it depends on the brightness difference between two objects, and for ground-based surveys, it is influenced by seeing. 

Considering catalogues rather than images, additional considerations must be taken into account. Close sources may have less accurate astrometry and photometry as a result of the disturbing presence of the other nearby source and may be preferentially filtered out from catalogue releases.
In addition, when the sky is observed at several different epochs or when fields of view overlap, different observations of the same source may not be recognised as such and duplicated entries are introduced in the catalogue.

The effective angular resolution of the external catalogue is important in order to recognise and correctly match the sources that are resolved in \emph{Gaia} but not in the external
catalogue (see Subsection~\ref{subsec:resultangres}). The comparison between the effective angular resolution values reported in Table~\ref{table:ExtProp} and the separation distribution in the catalogues is important to evaluate their consistency. In addition, the analysis of the effective angular resolution allows flagging suspected duplicates in the external catalogue, which also can hamper the cross-match results.

\begin{figure*}
   \centering
   \includegraphics{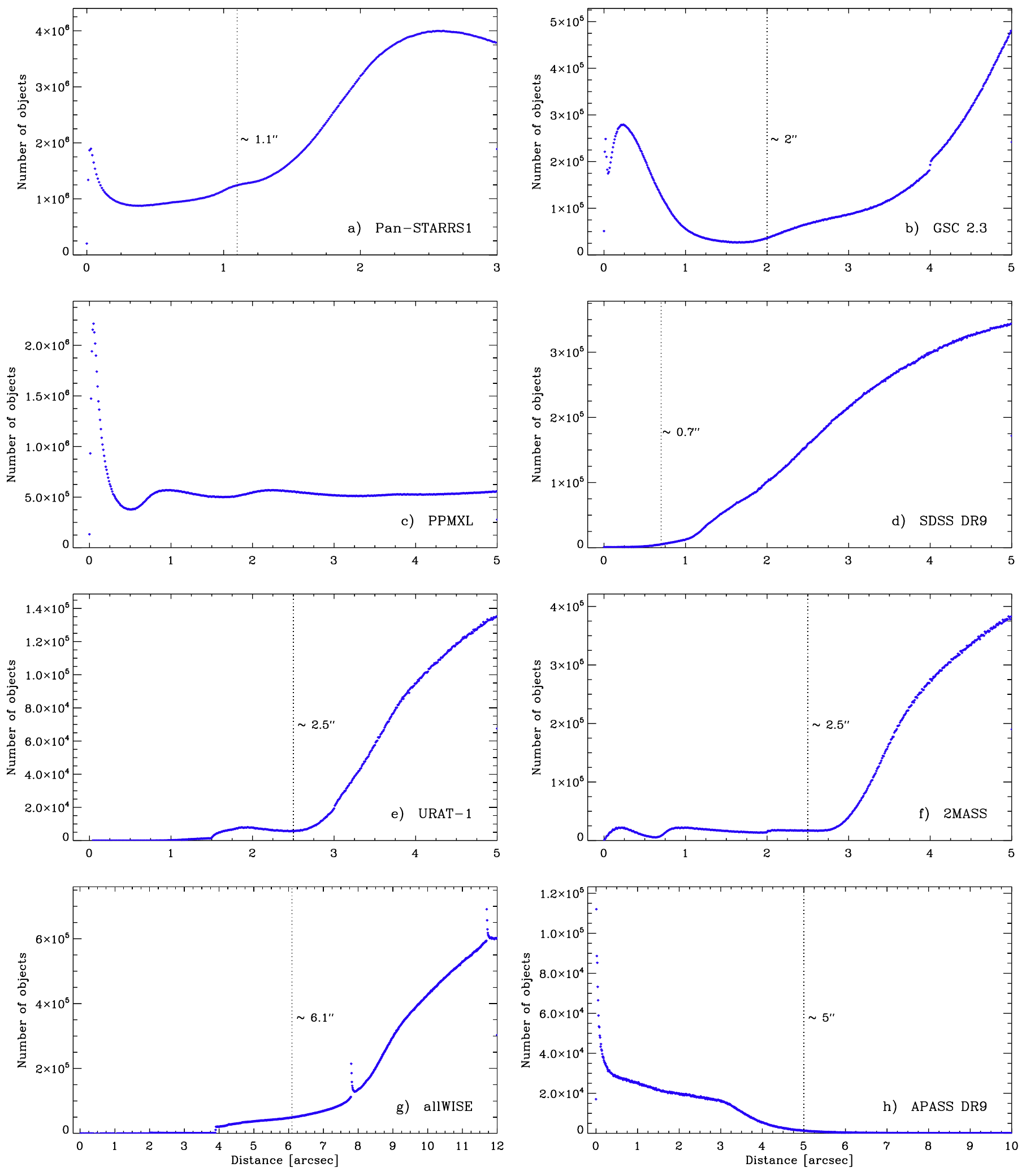}
   \caption{Angular distance distribution of nearest neighbours in each dense survey considered in this study. The vertical dotted lines indicate the angular resolution as defined in Table~\ref{table:ExtProp}.}
              \label{fig:FigAngResa}
    \end{figure*}

\begin{figure*}
   \centering
   \includegraphics[width=0.96\linewidth]{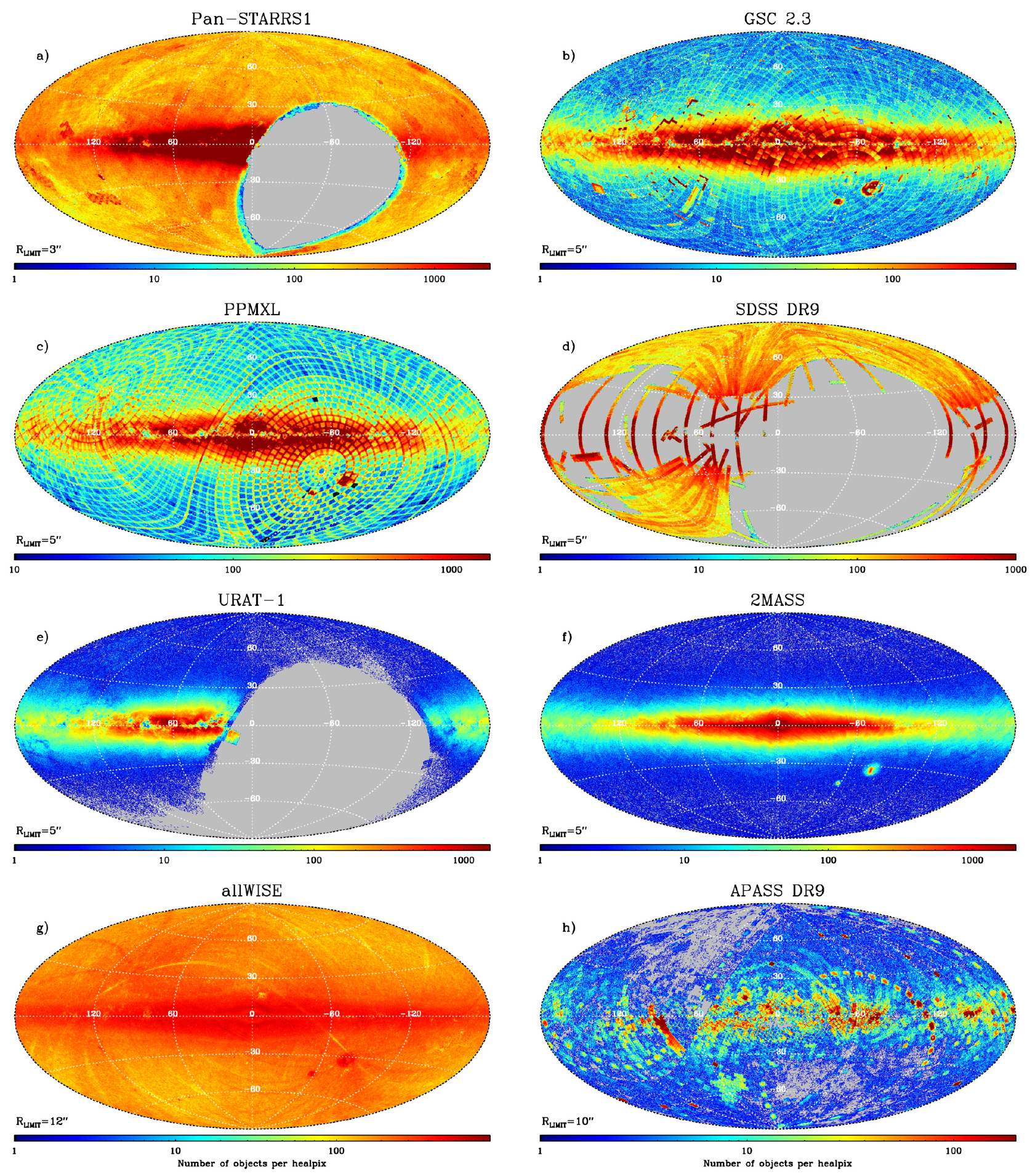}
   \caption{Sky maps that show for each dense survey the number of objects per healpix that have at least one neighbour within the radius indicated by $R_{\mathrm{LIMIT}}$. The maps are obtained with a HEALPix tessellation with resolution $N_{\mathrm{side}}=2^{8}$.}
              \label{fig:FigAngResb}
    \end{figure*}

In Figure~\ref{fig:FigAngResa} we show the results of a search of the nearest neighbour (neglecting additional neighbours except for the nearest) around each object in a given dense survey. We used a fixed radius of 5 arcsec for most surveys, with the exception of Pan-STARRS1~DR1, for which, given its size, we used a 3 arcsec radius, and allWISE and APASS~DR9, for which, given their resolution, we used 12 and 10 arcsec, respectively.
Figure~\ref{fig:FigAngResb} shows instead density maps (obtained using a HEALPix tessellation with resolution $N_{\mathrm{side}}=2^{8}$) of the number of sources with a nearest neighbour within the search radius defined above. 

The histograms in Figure~\ref{fig:FigAngResa} show the real distribution of nearest neighbours (i.e. the source separation distribution) in the catalogues and should be compared with the effective angular resolutions (dotted vertical lines) listed in Table~\ref{table:ExtProp}. 
Nearest neighbours much closer than the marked angular resolution are most probably duplicated
sources, while nearest neighbours at distances smaller than but close to the marked angular resolution are
still possibly truly distinct sources. 
The maps shown in Figure~\ref{fig:FigAngResb} allow evaluating whether the sky distribution of sources with a close nearest neighbour is correlated with known Galactic features or if they are instead related to the survey observation methods.

In Figure~\ref{fig:FigAngResa}, the expected histogram shape of a well-cleaned catalogue is similar to the shapes of SDSS DR9, URAT-1, 
allWISE, or 2MASS. 
The initial rise in the Pan-STARRS1~DR1 distance distribution indicates duplicated sources. In Figure~\ref{fig:FigAngResb} the fields observed by Pan-STARRS1~DR1 with higher source counts are clearly visible, together with some issues at the borders of the hexagonal gigapixel camera tiles. The over-densities are also clearly distinguishable in Figure~\ref{figure:all}.
The very small peak, visible at distances close to zero, in GSC~2.3 indicates the Tycho-2 and SKY2000 (\citealt{sky}) sources. These sources, which were added to complement GSC~2.3 at the bright end, cannot be considered duplicates. Conversely,  the second GSC~2.3 peak (around 0.25 arcsec) indicates the presence of duplicated sources that are mainly present at tile edges (see Figure~\ref{fig:FigAngResb}).
For PPMXL no analysis (or flagging of suspected duplicates) is possible because of the composite nature of the catalogue and because the original observed coordinates are not present in the catalogue, which includes only a set of positions propagated to J2000.0.
Nevertheless, it is clear from the map in Figure~\ref{fig:FigAngResb}, but also from Figure~2, panel d) in Paper~I, that there is a relevant issue with duplicates.
The reverse shape of  the APASS DR9 histogram implies that the catalogue is largely affected by duplicated sources: $\sim$14.3\% of the objects have at least one neighbour that is located too closely. The map in Figure~\ref{fig:FigAngResb} also shows several issues in completeness and duplicates.
 
This type of analysis will be the base for the planned further cross-match developments when we will deal with the possibly duplicated sources, and we will address the issues related to different angular resolutions between \emph{Gaia} and the external catalogues in a more complete way.

\end{appendix}
\end{document}